\documentclass[twocolumn]{aastex631}

\usepackage{CJK}
\usepackage{stackengine}

\newcommand{\xmm}{{\it XMM-Newton }}
\newcommand{\rej}{{RE~J1034+396 }}

\shorttitle{WAs of RE~J1034}
\shortauthors{Zhou et al.}

\begin{document}
\begin{CJK*}{UTF8}{gbsn}

\title{On the Connection between the Repeated X-ray Quasi-periodic Oscillation and Warm Absorber in the Active Galaxy RE~J1034+396}

\correspondingauthor{Taotao Fang, Junjie Mao}
\email{fangt@xmu.edu.cn}
\email{jmao@tsinghua.edu.cn}

\author[0009-0000-3060-1219]{Zheng Zhou (周正)}
\affiliation{Department of Astronomy, Xiamen University, Xiamen, Fujian 361005, China}

\author[0000-0001-7557-9713]{Junjie Mao (毛俊捷)}
\affiliation{Department of Astronomy, Tsinghua University, Haidian DS 100084, Beijing, China}
\affiliation{SRON Netherlands Institute for Space Research, Sorbonnelaan 2, 3584 CA Utrecht, the Netherlands}

\author[0000-0002-2853-3808]{Taotao Fang (方陶陶)}
\affiliation{Department of Astronomy, Xiamen University, Xiamen, Fujian 361005, China}

\author[0000-0002-1010-7763]{Yijun Wang (王倚君)}
\affiliation{Department of Astronomy, Nanjing University, Nanjing 210093, China}
\affiliation{Key Laboratory of Modern Astronomy and Astrophysics (Nanjing University), Ministry of Education, Nanjing 210093, China}

\author[0000-0002-6896-1364]{Fabrizio Nicastro}
\affiliation{Istituto Nazionale di Astrofisica (INAF) - Osservatorio Astronomico di Roma Via Frascati 33 00078 Monte Porzio Catone (RM), Italy}
\affiliation{Department of Astronomy, Xiamen University, Xiamen, Fujian 361005, China}

\author[0000-0001-7829-7797]{Jiayi Chen (陈佳怡)}
\affiliation{Department of Astronomy, Tsinghua University, Haidian DS 100084, Beijing, China}

\begin{abstract}

We conduct an in-depth spectral analysis of $\sim1$~Ms \xmm data of the narrow line Seyfert~1 galaxy RE~J1034+396. The long exposure ensures high spectral quality and provides us with a detailed look at the intrinsic absorption and emission features toward this target. Two warm-absorber (WA) components with different ionization states ($\log (\xi/{\rm erg~cm~s}^{-1}) \sim 4$ and $\log (\xi/{\rm erg~cm~s}^{-1}) \sim 2.5-3$) are required to explain the intrinsic absorption features in the RGS spectra. The estimated outflow velocities are around $-1400$~km~s$^{-1}$ and $-(100-300)$~km~s$^{-1}$ for the high- and low-ionization WA components, respectively. Both absorbers are located beyond the broad-line region and cannot significantly affect the host environment. We analyze the warm absorbers in different flux states. We also examine the May-2007 observation in the low and high phases of quasi-periodic oscillation (QPO). In contrast to previous analyses showing a negative correlation between the high-ionization WA and the QPO phase, we have found no such variation in this WA component. We discover a broad emission bump in the spectral range of $\sim12-18$~\AA, covering the primary features of the high-ionization WA. This emission bump shows a dramatic change in different source states, and its intensity may positively correlate with the QPO phase. The absence of this emission bump in previous work may contribute to the suggested WA-QPO connection.

\end{abstract}

\keywords{Seyfert galaxies (1447) --- X-ray astronomy (1810) --- Warm ionized medium (1788) --- High resolution spectroscopy (2096)}

\section{Introduction} \label{sec:intro}

Outflowing gas from Active Galactic Nuclei (AGNs) acts as a bridge that connects the central black hole (BH) and the host galaxy. It influences the surroundings of the BH and may regulate the environments of the host galaxy (see, e.g., \citealp{King2015}). Characterizing the physical properties of these outflowing winds is crucial for understanding their origins and getting knowledge on the growth of BH and its coevolution with the host (e.g., \citealp{Kormendy2013}).

Warm absorbers (WAs) are a type of AGN outflow discovered through absorption lines and edges in the soft X-rays (e.g., \citealp{Halpern1984, Nicastro1999, Blustin2005, Steenbrugge2005, Krongold2007, McKernan2007, Laha2014}). It has been reported that the WAs can be detected in around $50-65\%$ of the nearby type~1 AGNs \citep{Blustin2005, McKernan2007, Tombesi2013, Laha2014}. Considering a typical lifetime of an AGN of $10^7$ years, the high detection rate implies WAs to be long-lived and highly-covering outflowing clouds. The kinetic energy of the WAs is usually insufficient to generate significant feedback (e.g., \citealp{Krongold2007, Reeves2009, Ebrero2011, Gupta2013}). However, WAs may still impact significantly the host surrounding environment during the entire lifetime \citep{Blustin2005, Khanna2016}.

WAs are usually found to be multi-phase, multi-component winds (e.g., NGC~3783, \citealp{Kaspi2002, Krongold2003, Mao2019}; NGC~4051, \citealp{Krongold2007}; NGC~5548, \citealp{Kaastra2000, Steenbrugge2003, Steenbrugge2005}; IRAS~13349+2438, \citealp{Sako2001}; MR~2251-178, \citealp{Reeves2013}). Spectral features of these WA components could cover a wide range of ionization states from neutral species, such as the unresolved transition array of inner-shell Fe \citep{Sako2001, Behar2001}, to highly ionized H- or He-like ions (e.g., \citealp{Reynolds1997}). Typical WA winds manifest outflow velocities of $v_{\rm out} <2000$~km~s$^{-1}$ and ionization parameters\footnote{Here we adopt the definition of ionization parameter following \cite{Tarter1969} as $\xi = L_{\rm ion}/(n_{\rm H}r^2)$, where $L_{\rm ion}$ is the $1-1000$~Ryd source luminosity, $n_{\rm H}$ the gas density, and $r$ the radial distance between the wind and the central illuminating source.} of $\log (\xi/{\rm erg~cm~s}^{-1}) <4$ (e.g., \citealp{McKernan2007, Laha2021}).

The study of WAs remains incomplete ever since the first report of this outflow phenomenon \citep{Halpern1984}. An important debate is on the launching mechanism, with several theoretical models proposed to explain wind generation. The widest accepted ones describe the WAs as thermally-evaporated winds (e.g., \citealp{Krolik2001, Dorodnitsyn2008, Mizumoto2019}), radiatively-driven winds (e.g., \citealp{Proga2004, Dannen2019}), or magneto-driven winds (e.g., \citealp{Blandford1982, Fukumura2010, Fukumura2018}). We also know very little about the wind location. The most frequently adopted method calculates the radial distance $r$ through the definition of the ionization parameter. However, the degeneracy between $n_{\rm H}$ and $r$ makes the distance estimation ambiguous.

There are two approaches to determining the density of an outflowing absorber and thus breaking the $n_{\rm H}-r$ degeneracy. The first is to analyze the response of the absorber to the source flux change and determine the recombination timescale of the wind plasma (e.g., \citealp{Krongold2005, Krongold2007, Kaastra2012, Khanna2016, Wang2022}), which inversely correlates with the gas density \citep{Nicastro1999, Bottorff2000}. The second approach is through analysis of the metastable absorption lines that are sensitive to the plasma density (e.g., \citealp{Kaastra2004, Mao2017}). However, accurately determining the plasma density using either method is challenging. Measuring the recombination timescale requires an apparent change in the AGN luminosity, long exposures, and a high plasma density to ensure the recombination timescale is detectable (e.g., \citealp{Krongold2007, Kaastra2012, Silva2016, Rogantini2022}). As for the analysis of metastable lines, the insufficient effective area and spectral resolution of the current generation X-ray spectrometers make the lines hard to detect.


RE~J1034+396 is a narrow line Seyfert~1 galaxy at redshift $z\sim0.0431$ and is famous for a repeated one-hour signal of quasi-periodic oscillation (QPO) \citep{Gierliski2008}. The BH mass of this target is between $\sim10^6-10^7~M_\odot$ (\citealp{Czerny2016, Bian2010, Chaudhury2018, Jin2021}). In this work, we adopt a recently estimated mass of $M_{\rm BH} = 3\times10^6~M_\odot$ \citep{Chaudhury2018, Jin2021}.

By analyzing the \xmm EPIC-pn data observed in May 2007, \cite{Maitra2010} identified an O\,{\sc viii} absorption edge at $\sim0.85-1.10$~keV, presenting the first evidence for WA in this target. The hydrogen column density and the ionization parameter of the WA constrained by the oxygen edge were $N_{\rm H}=(4\pm1)\times10^{21}$~cm$^{-2}$ and $\log (\xi/{\rm erg~cm~s}^{-1}) = 3.2^{+0.2}_{-0.1}$, respectively. The authors also found that the oxygen edge disappeared when the QPO reached the high phase, suggesting the QPO is produced by the periodic obscuring of the WA clouds located at $15$~$R_{\rm g}$, where $R_{\rm g}$ is the gravitational radius. The WA was later confirmed by the narrow absorption lines of O\,{\sc viii}, Fe\,{\sc xviii}, Fe\,{\sc xix}, etc, seen in the \xmm RGS spectrum \citep{Middleton2011}, with $N_{\rm H}=2.23\times10^{21}$~cm$^{-2}$, $\log (\xi/{\rm erg~cm~s}^{-1}) = 2.7$, and an outflow velocity on the order of $1000$~km~s$^{-1}$. This finding casts doubt on the obscuring scenario since the location of $15$~$R_{\rm g}$ is too small to explain the narrow broadening of the lines ($\sim400-1000$~km~s$^{-2}$) seen in the high-resolution spectrum. \cite{Jin2021} studied the May-2007 and Oct-2018 spectra by a similar method applied in \cite{Maitra2010} and found the equivalent width of the most significant WA feature (Fe\,{\sc xix} at $\sim0.9$~keV) negatively correlates with the QPO phase. Therefore, they concluded that the WA is possibly correlated with the QPO phase.

RE~J1034+396 is the only AGN for which a correlation between the WA and QPO is hinted. The potential existence of this correlation may open up a new avenue to study the nature of both the phenomena and the vicinity of the central BH. However, no detailed analysis of the WAs in this target has been presented. With more than $1$~Ms \xmm observations over the past dozen years, the high spectral quality makes RE~J1034+396 one of the best targets to study the WA properties.

In this work, we aim to constrain the basic properties of WAs in RE~J1034+396. We also inspect the spectral time variation at around $0.9$~keV, which will benefit our understanding of the potential connection between WA and QPO. This paper is organized as follows. In section~\ref{sec:obs_data}, we introduce the \xmm observations and the data reduction procedures. In section~\ref{sec:line_analysis}, we describe the spectral fitting methods. The fitting result and discussions are presented in section~\ref{sec:result_discuss}. Throughout the paper, we adopt a flat ${\rm \Lambda CDM}$ model for all the calculations, with $H_0 = 70$~km~s$^{-1}$~Mpc$^{-1}$, $\Omega_\Lambda = 0.70$, and $\Omega_{\rm M} = 0.30$.

\section{Observation and data resuction} \label{sec:obs_data}

\begin{deluxetable*}{ccccccccc}
\tablewidth{0pt}
\setlength{\tabcolsep}{1.7 mm}
\tablecaption{\xmm observation log. The first two columns are the observational date and ID. The total and net exposures of EPIC-pn and RGS after filtering out the soft protons are listed in columns~(3) to (5). Column~(6) shows the $0.3-10.0$~keV source count rates (correlated by the background) observed by the EPIC-pn. Columns~(7) and (8) are the available OM filter and the $UVW1$ count rates. The last column labels the X-ray flux state of each observation. At the bottom of the table, we show the exposure of each combined spectrum.
\label{tab:obs_log}}
\tablewidth{400pt}
\tablehead{ 
	\colhead{Start time} & \colhead{obs. ID} & \colhead{Duration} & RGS net & EPIC-pn net & \colhead{pn count rates} & \colhead{OM filter} & $UVW1$ count rates & \colhead{State}  \\
	\colhead{} & \colhead{} & \colhead{(ks)} & \colhead{Exp. (ks)} & \colhead{Exp. (ks)} & \colhead{(counts~s$^{-1}$)} & & \colhead{(counts~s$^{-1}$)}
}
\decimalcolnumbers
\startdata
2009-05-31 & 0561580201 & $70.3$ & $56.8$ & $33.6$ & $5.65$ & $UVW1$ & $3.805\pm0.011$ & HFS \\
2010-05-09 & 0655310101 & $52.2$ & $24.6$ & $17.1$ & $4.13$ & $UVW1$ & $3.771\pm0.011$ & LFS \\
2010-05-11 & 0655310201 & $54.1$ & $36.9$ & $25.8$ & $4.01$ & $UVW1$ & $3.795\pm0.011$ & LFS \\
2011-05-07 & 0675440101 & $37.4$ & $21.0$ & $14.2$ & $4.32$ & $UVW1$ & $3.805\pm0.013$ & LFS \\
2011-05-27 & 0675440201 & $37.8$ & $16.9$ & $10.3$ & $4.03$ & $UVW1$ & $3.728\pm0.014$ & LFS \\
2011-05-31 & 0675440301 & $37.0$ & $20.5$ & $13.9$ & $6.96$ & $UVW1$ & $3.869\pm0.013$ & HFS \\
2018-10-30 & 0824030101 & $73.5$ & $70.9$ & $49.2$ & $4.00$ & $UVW1$ & $3.792\pm0.009$ & LFS \\
2020-11-20 & 0865010101 & $90.0$ & $70.4$ & $45.5$ & $4.09$ & $V$ & \nodata & LFS \\
2020-12-01 & 0865011001 & $87.0$ & $82.4$ & $56.0$ & $4.28$ & $U$ & \nodata & LFS \\ 
2020-12-03 & 0865011201 & $92.9$ & $89.1$ & $58.6$ & $4.14$ & $UVW1$ & $4.204\pm0.009$ & LFS \\
2020-12-05 & 0865011101 & $91.0$ & $87.7$ & $61.3$ & $4.35$ & $B$ & \nodata & LFS \\
2021-04-24 & 0865011301 & $94.0$ & $90.3$ & $61.7$ & $3.93$ & $UVM2$ & \nodata & LFS \\
2021-05-02 & 0865011401 & $89.0$ & $86.5$ & $59.9$ & $4.18$ & $UVW2$ & \nodata & LFS \\
2021-05-08 & 0865011501 & $93.0$ & $90.1$ & $61.9$ & $4.08$ & $UVW1$ & $4.193\pm0.009$ & LFS \\
2021-05-12 & 0865011601 & $92.0$ & $82.5$ & $53.1$ & $4.16$ & $UVW1$ & $4.183\pm0.010$ & LFS \\
2021-05-16 & 0865011701 & $94.0$ & $87.2$ & $54.4$ & $4.18$ & $UVM2$ & \nodata & LFS \\
2021-05-31 & 0865011801 & $94.0$ & $87.0$ & $57.9$ & $3.92$ & $UVW2$ & \nodata & LFS \\
\hline
Stacked LFS & \nodata & $1171.9$ & $1023.5$ & $686.9$ & \nodata & \nodata & \nodata & \nodata \\
Stacked HFS & \nodata & $107.3$ & $77.3$ & $47.5$ & \nodata & \nodata & \nodata & \nodata \\
\enddata
\end{deluxetable*}


\rej was observed by \xmm for more than $1$~Ms before 2021 May 31. We use the X-ray data of both the EPIC-pn and RGS observed in the small-window mode. This yielded $17$ available observations, and Table~\ref{tab:obs_log} shows the detailed information of each observation.
To accurately estimate the ionizing luminosity $L_{\rm ion}$, we also used the OM data in spectral fitting to constrain the broadband spectral energy distribution (SED). We only considered the {\it UVW1} data of OM since the target was primarily observed by this filter, and the optical emission of \rej is dominated by the starlight of the host galaxy \citep{Bian2010, Czerny2016, Jin2021}.

We followed the standard procedures of the \xmm Science Analysis System ({SAS} v21.0.0) to reduce all the data. The EPIC-pn data were processed using the task {\it epproc}. We filtered out bad pixels by setting `${\rm FLAG}=0$' and only adopted the single events (`${\rm PATTERN} = 0$') to reduce the pile-up effect. Source spectra were extracted by a $30\arcsec$ circular region centered on the target, and the background spectra were taken from a $30\arcsec$ source-free circle of the same chip. The $10.0-12.0$~keV light curve was applied to filter the soft-proton flares with a threshold of ${\rm RATE}<0.04$~counts~s$^{-1}$. As for the RGS, we used the task {\it rgsproc} to reduce the data. The light curve of CCD~9 of each RGS unit was adopted to exclude the bad intervals with rates above $0.1$~counts~s$^{-1}$. Only the first-order spectra were considered. Similar to \cite{Mehdipour2015}, we processed the OM data by the standard task {\it omichain}. The derived {\it UVW1} count rates in the source list were later converted into the standard OGIP format through the task {\it om2pha}.

\begin{figure}
\includegraphics[width=0.47\textwidth]{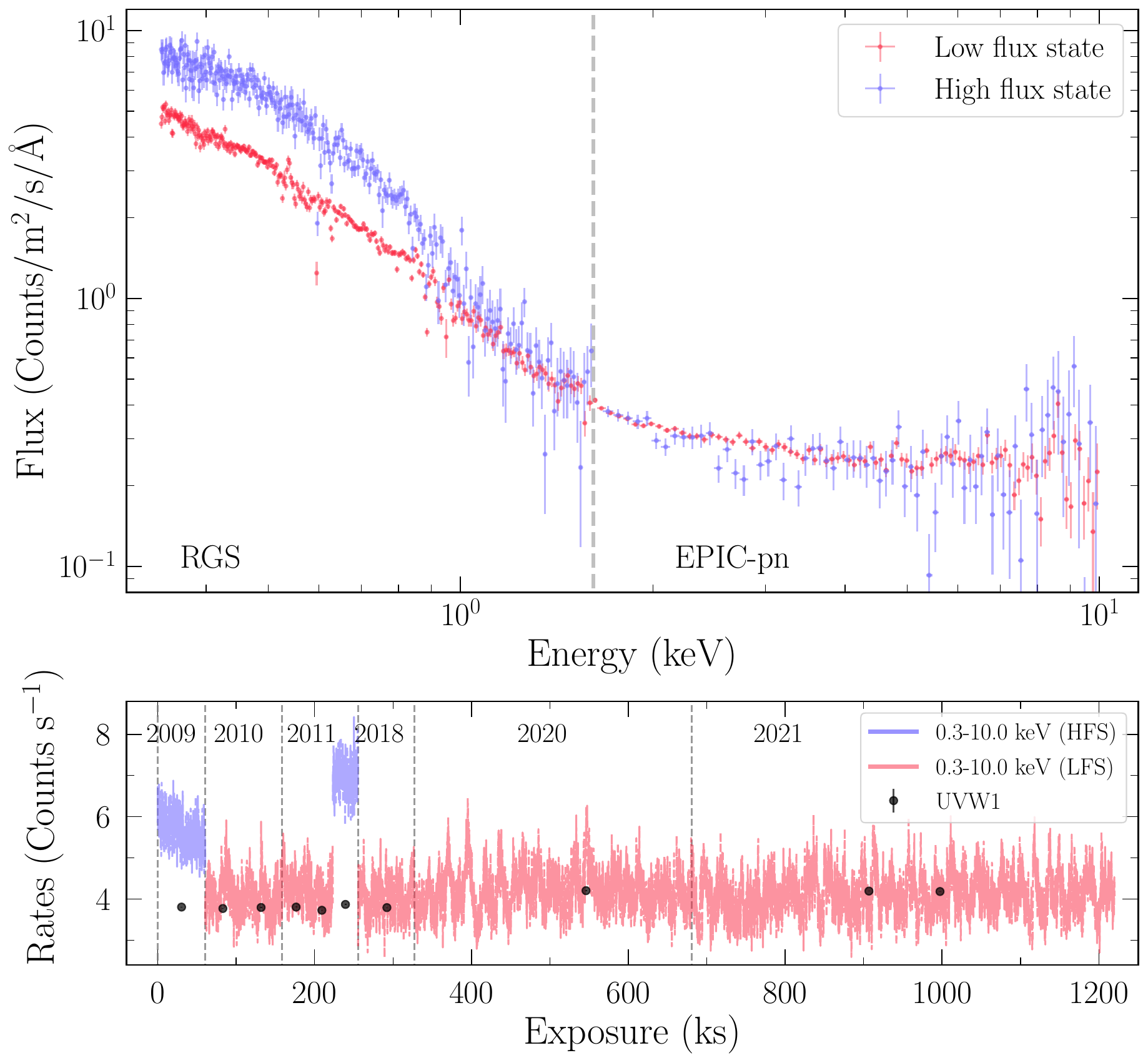}
\caption{(Top) Co-added $0.3-10.0$~keV X-ray spectra of the low flux state (red) and the high flux state (blue). The left side of the vertical line shows the RGS data, and the right side is the EPIC-pn data. We make an instrumental calibration between RGS and EPIC-pn, and the EPIC-pn spectra are re-scaled by a factor of $0.922$ and $1.054$ for the LFS and HFS, respectively. We re-bin the spectra for display purposes. (Bottom) Red and blue show the LFS and HFS light curves at $0.3-10$~keV band overlapped with the UVW1 count rates in black dots.} 
\label{fig:pn_continuum}
\end{figure}

Over the $12$ years, the target shows two X-ray flux states with a maximum flux change of $\sim50\%$ at $0.3-10.0$~keV spectral range. It has been reported that the QPO of \rej can only be detected at the low flux state (\citealp{Zoghbi2011, Alston2014}). We, therefore, divided the spectra into two catalogs to track the WA properties in different flux states (see Figure~\ref{fig:pn_continuum}). We also provide $1-4$~keV light curves of the May-2007 and Oct-2018 observations for illustration purposes of the QPO in Appendix~\ref{sec:LC_PSD}. The high-flux-state (HFS) spectrum consists of the two brightest observations (obs.ID 0561580201 and 0675440301) with a total net exposure of $77$~ks and $48$~ks for RGS and EPIC-pn, respectively. The other observations make up the low-flux-state (LFS) spectrum with a total net exposure of $1024$~ks for RGS and $687$~ks for EPIC-pn. To analyze the weak features with a high signal-to-noise ratio, we co-added all the spectra in each catalog using the tools {\it epicspeccombine} and {\it rgscombine} for EPIC-pn and RGS, respectively. Signal-to-noise ratio per resolution element at $20$~\AA\ is $26$ and $11$ for the stacked LFS and HFS RGS spectra, respectively.
Despite the X-ray flux state, the {\it UVW1} count rates show a small variation ($\lesssim11\%$) among the $10$ observations when the filter is available. The amplitude of the UV variation is comparative to the $0.3-10$~keV flux change of $\sim10\%$ among the LFS observations. Thus, all the {\it UVW1} data were combined using $1/\sigma^2$ as the weight for spectral fitting of both the flux states, where $\sigma$ is the uncertainty of the {\it UVW1} count rates.

The EPIC-pn, RGS, and OM spectra were analyzed together. We cross-calibrated the EPIC-pn and RGS to match their flux at a common wavelength band. The EPIC-pn flux was later rescaled by a factor of $0.922$ and $1.054$ for the co-added LFS and HFS spectra, respectively. In Figure~\ref{fig:pn_continuum}, we show the stacked LFS and HFS spectra of RGS and EPIC-pn. The EPIC-pn spectra exhibit some wiggles at $\gtrsim7$~keV spectral range, implying the presence of ultra-fast outflow (UFO, see Appendix~\ref{sec:ufo_search}). The source has a very soft spectrum, and the variability between the two states is mainly due to the changes in the soft X-rays.

\section{Spectral Analysis} \label{sec:line_analysis}

In this paper, we use the SPEX package v3.07.03 \citep{Kaastra1996, Kaastra2020} to fit the time-averaged spectra and search for the optimal model by minimizing the C-statistic \citep{Kaastra2017}. Solar abundance \citep{Lodders2009} is adopted throughout the paper unless otherwise mentioned. Uncertainties are quoted at $1\,\sigma$ significance range.
To avoid oversampling the data, we re-binned the RGS spectra by a factor of three. The EPIC-pn data were re-binned optimally by the SPEX command {\it obin} \citep{Kaastra2016}.

\subsection{Stacked Low-flux-state Spectrum}
\subsubsection{Continuum Modelling}


The intrinsic SED consists of a disk blackbody component ({\it dbb}), a warm Comptonized disk component ({\it comt}), and a power-law tail ({\it pow}) \citep{Middleton2010, Hu2014, Jin2021}. 
The optical depth of the Comptonized disk was fixed at $\tau=11$ according to previous works \citep{Done2012, Hu2014, Kaufman2017, Jin2021}. The seed photons of the warm Comptionization were assumed to come from the inner disk emission (i.e., $T_{dbb}$ was linked to $T_{0}$ of {\it comt}). We adopted the plasma temperature of the Comptonized disk to constrain the low-energy cut-off for the power-law component, and the high-energy cut-off was set to be $300$~keV \citep{Gonzalez2018, Buhariwalla2020}. We modeled the Galactic neutral absorption with the {\it hot} model \citep{de_Plaa2004, Steenbrugge2005}. This model accounts for both absorption lines and edges under collisional ionization equilibrium. We let the gas temperature free to vary but fixed the hydrogen column density at $1.25\times10^{20}$~cm$^{-2}$ according to the HI4PI survey \citep{HI4PI}. The best-fit continuum model gives $C=2293$, with $1049$ degrees of freedom (DoF). The {\it dbb} component was fixed at the best-fit value for further analysis.

The neutral gas of the host galaxy was also considered at the beginning by adding another {\it hot} component at the source redshift. However, the improvement of the fitting is insignificant ($\Delta C \sim 0$), and we did not find any neutral absorption lines like O\,{\sc i}, O\,{\sc ii}, and N\,{\sc i} at the host redshift. The best-fit hydrogen column density is $N_{\rm H,host} < 4\times10^{18}$~cm$^{-2}$. This value is much smaller than the previous results of $N_{\rm H,host} \sim (1-6)\times10^{20}$~cm$^{-2}$ when only absorption edges were considered (e.g. \citealp{Done2012,Jin2021}). Therefore, we excluded this component due to its negligible contribution.

\subsubsection{Absorption from Galactic Warm-hot Halo}

\begin{figure}
\includegraphics[width=0.47\textwidth]{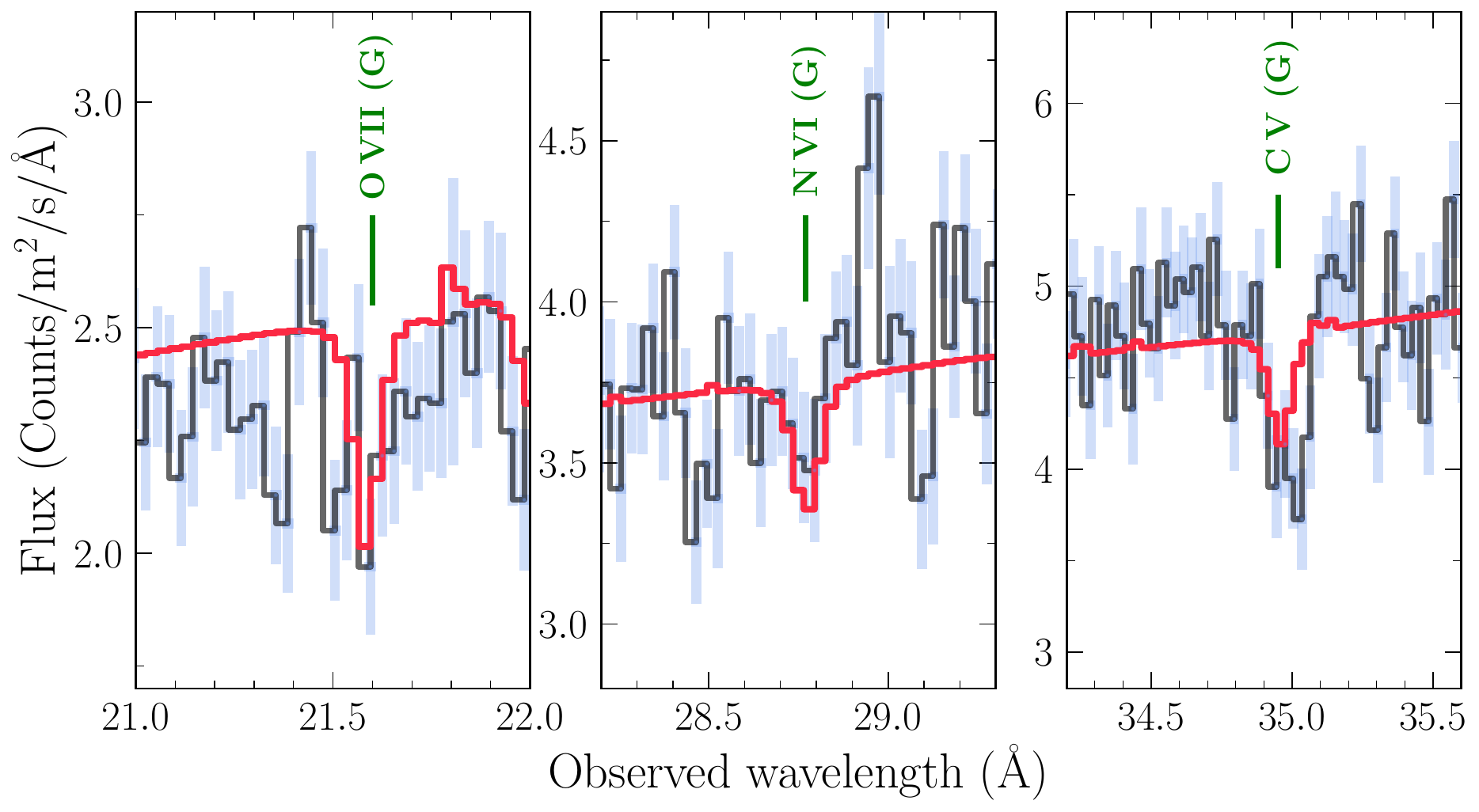}
\caption{Absorption lines from Galactic warm-hot halo. The black curve is the observed data with $1\,\sigma$ uncertainty shown in blue. The red curve is the best-fit {\it hot} model. The Galactic C\,{\sc v} ${\rm He}\beta$ line is overlapped with the AGN absorption features.}
\label{fig:Gal_halo}
\end{figure}

We identify three absorption features centered at $21.618$~\AA, $28.773$~\AA, and $34.946$~\AA\ (see Figure~\ref{fig:Gal_halo}), which are consistent with the rest wavelengths of O\,{\sc vii} ${\rm He}\alpha$, N\,{\sc vi} ${\rm He}\alpha$, and C\,{\sc v} ${\rm He}\beta$, respectively. These absorption lines come from the warm-hot halo (see, e.g., \citealp{Tumlinson2017}) of the Milky Way (MW) with a temperature of $10^5-10^7$~K. The ion fractions of all three species peak at $\sim10^5-10^6$~K under collisional ionization equilibrium \citep{Gnat2007}, implying the warm phase ($10^5-10^6$~K, e.g., \citealp{Sembach2003, Savage2009, Tumlinson2017, Qu2020}) primarily contributes to the lines. Another {\it hot} model was used to fit the Galactic warm-hot lines, with the gas temperature a free parameter varied between $10^5$~K and $10^{7}$~K. The best-fit result, as shown in Figure~\ref{fig:Gal_halo}, improves the fitting by $\Delta C / \Delta {\rm DoF} = 59/2$ ($C/{\rm DoF}=2234/1047$). The local C\,{\sc v} ${\rm He}\beta$ line is overlapped with the AGN intrinsic features.


\subsubsection{Warm Absorber}

We applied the photoionization model {\it pion} \citep{Miller2015, Mehdipour2016, Mao2018} to fit the features from the warm absorber. {\it Pion} is a self-consistent model which calculates the thermal equilibrium, ionization balance, and absorption and emission spectrum of the photoionized gas. All the {\it pion} components can be simultaneously fitted with the spectral SED and other model components. The ionization balance is re-calculated in real-time during each iteration when the SED varies. Therefore, there is no requirement for a fixed prior SED as in the classical photoionization models.

\begin{figure}
\includegraphics[width=0.47\textwidth]{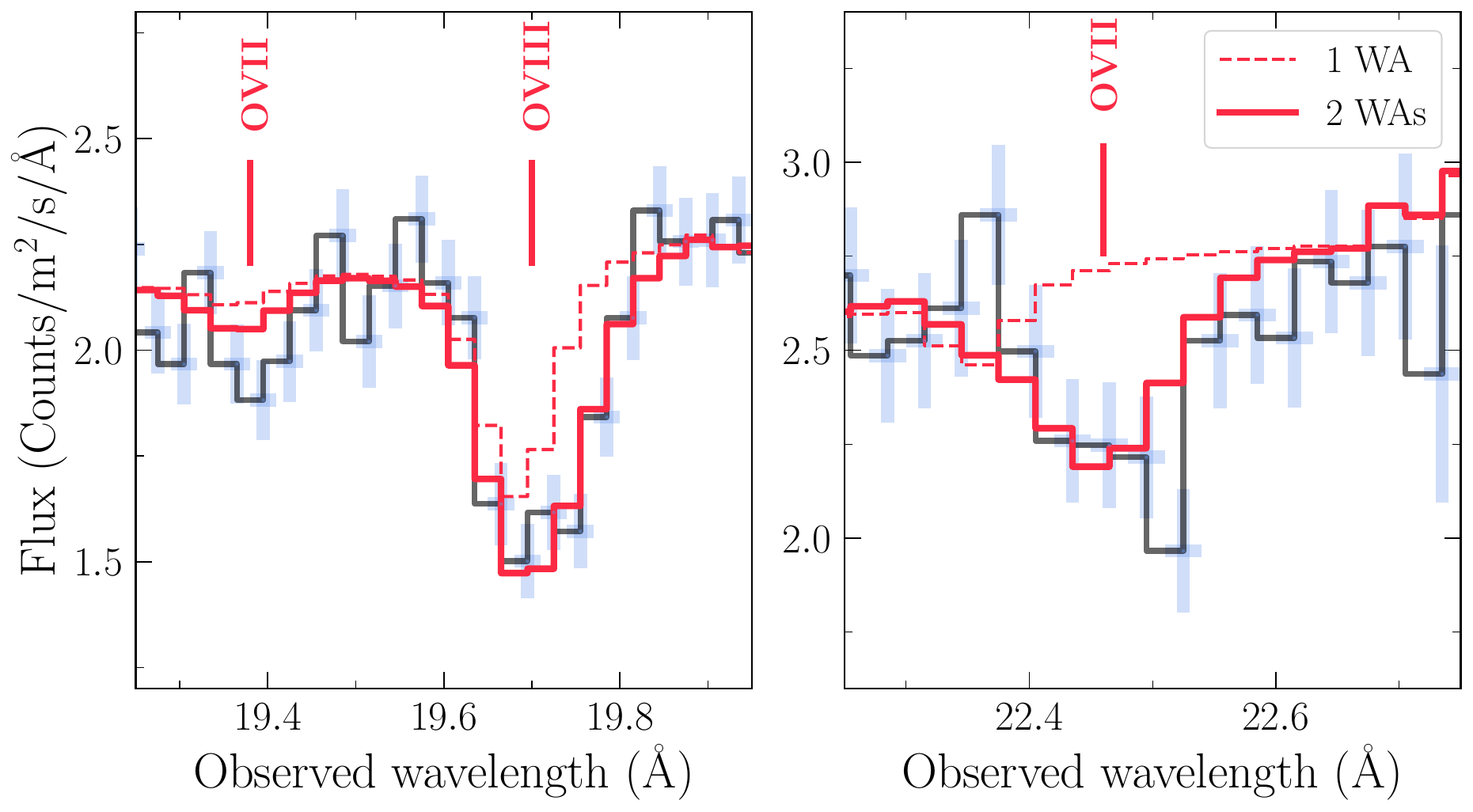}
\caption{Intrinsic O\,{\sc vii} and O\,{\sc viii} absorption lines in the co-added LFS spectrum. The black curve shows the observed spectrum with $1\,\sigma$ uncertainty in blue. The red-dashed and red-solid lines denote the best-fit modes with one and two WA components, respectively. 
}
\label{fig:1or2_WA}
\end{figure}

Initially, we added one {\it pion} component to fit the WA features and set the WA properties to the previous results \citep{Maitra2010, Middleton2011, Jin2021}: hydrogen column density of $N_{\rm H}=3\times10^{21}$~cm$^{-2}$, ionization parameter of $\log (\xi/{\rm erg~cm~s}^{-1}) = 3$, outflow velocity of $v_{\rm out} = -2000$~km~s$^{-1}$, and microscopic motion velocity (i.e., turbulence) of $v_{\rm mic} = 300$~km~s$^{-1}$. Fitting this component improves the best-fit model by $\Delta C / \Delta {\rm DoF} = 326 / 4$. This WA component (${\rm WA}_1$ hereafter) takes responsibility for the highly ionized Fe lines at around $\sim14$~\AA\ and part of the O\,{\sc viii} line at $19.7$~\AA\ (observer frame), which is similar to previous findings \citep{Middleton2011, Jin2021}. However, absorption features with a lower ionization are not well-modeled by ${\rm WA}_1$, especially for the intrinsic O\,{\sc vii} features shown in Figure~\ref{fig:1or2_WA}. We thus adopted the second WA component with a lower ionization state (${\rm WA}_2$ hereafter). This process improved the fitting by $\Delta C / \Delta {\rm DoF} = 120 / 4$ ($C / {\rm DoF} = 1788/1039$). In our modeling, the covering factor of both the WA components was assumed to be unity (full covering). The unobscured SED was adopted to ionize the ${\rm WA}_1$, while the leaked light from the ${\rm WA}_1$ layer was assumed to be the ionizing SED for ${\rm WA}_2$.

The two WA components help us recognize all the prominent absorption features in the LFS spectrum. However, some of the absorption features are not well described by the current model, implying additional absorption components. We regarded the third absorption component either as a WA modeled by a {\it pion} component or as the warm-hot halo of the host galaxy modeled by a {\it hot} component. However, the improvement of the fitting is limited in both cases. We found some emission features in the LFS spectrum overlap with absorption lines. Modeling these emission features may alleviate the imperfect fitting of the absorption lines.

\subsubsection{Emission Features}

\begin{figure}
\includegraphics[width=0.47\textwidth]{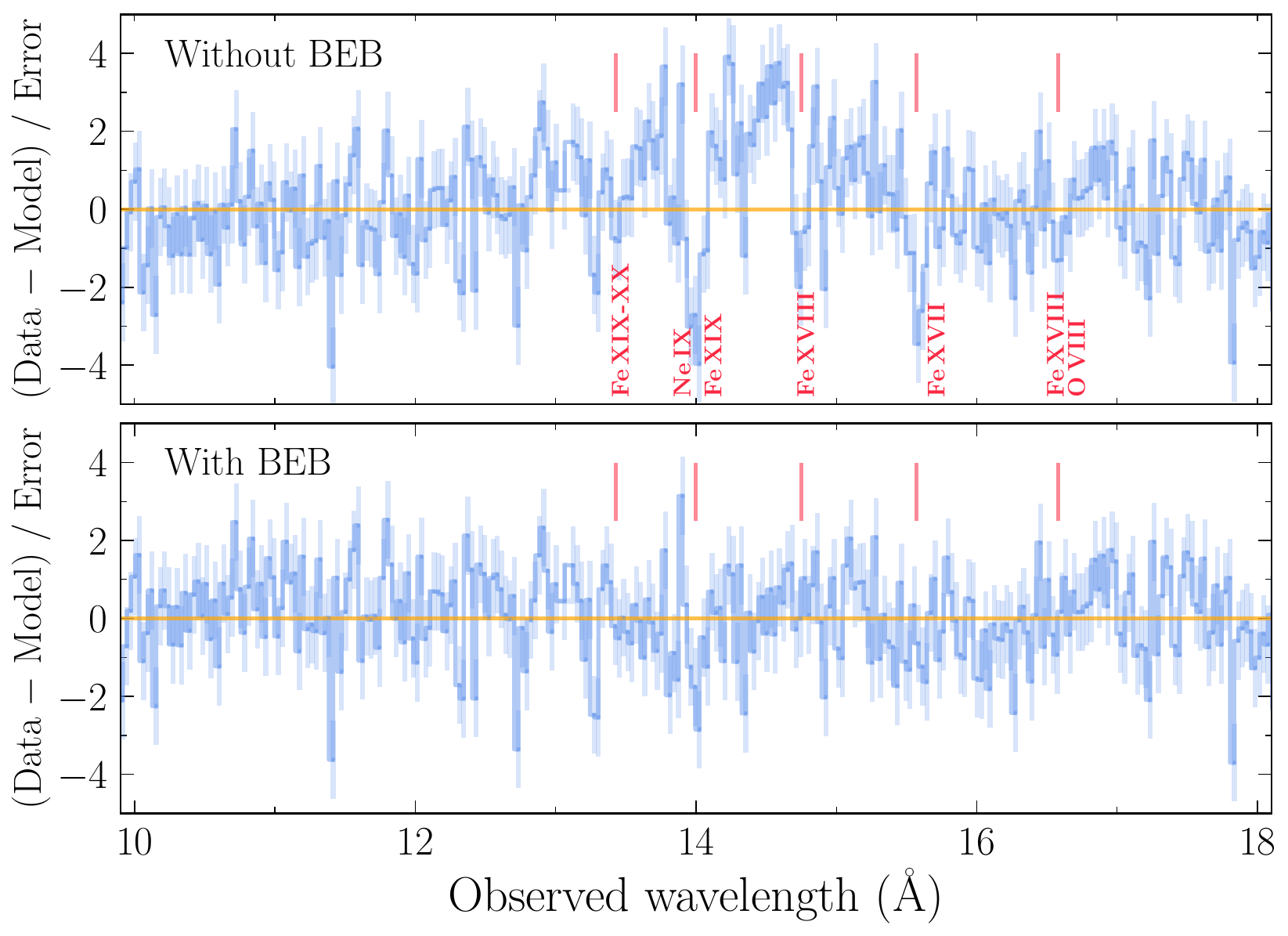}
\caption{Spectral residual without modeling the BEB (top) and modeling the BEB by a broad Gaussian component (bottom). The main absorption features of ${\rm WA}_1$ are labeled in red. Residuals of the absorption features disappear when BEB is introduced. }
\label{fig:resi_BEB}
\end{figure}

The most significant emission feature in the residual spectrum is a broad emission bump (BEB) at around $12-16$~\AA\ as shown in the top panel of Figure~\ref{fig:resi_BEB}. The BEB covers the primary features of ${\rm WA}_1$ and influences the accurate estimation of the WA property. We find the BEB can be well-modeled by a broad Gaussian profile as described by the {\it gaus} model in SPEX. The best-fit statistic of applying a broad Gaussian is $C / {\rm DoF} = 1489/1036$, with $\Delta C / \Delta {\rm DoF} = 299 / 3$. Residual spectrum after adding the broad Gaussian is shown in the bottom panel of Figure~\ref{fig:resi_BEB}. Absorption features of ${\rm WA}_1$, like Ne\,{\sc ix} and Fe\,{\sc xvii-xix} at around $14$~\AA, can then be well-modeled after introducing the broad Gaussian component. The BEB is hard to be explained by a photoionized emitting plasma (see Appendix~\ref{sec:BEB_pion}). Since the primary objective of this paper is to analyze the WA property, we applied this phenomenological model for further analysis.

Finally, we included a {\it pion} emission component to model the narrow emission lines from the warm emitter (WE). The unabsorbed SED is assumed to ionize the emission {\it pion} component \citep{Mao2018}. The velocity shift of the narrow emission lines is fairly slow, and we thus fixed the outflow velocity at $0$ to reduce the fitting complexity. We introduced the macroscopic motion broadening to the emission {\it pion} component modeled by a Gaussian broadening model {\it vgau}. This broadening refers to the rotation around the black hole and is often degenerated with the microscopic motion velocity \citep{Mao2018}. We fixed the microscopic velocity at $v_{\rm mic} = 100$~km~s$^{-1}$ and let the emission covering factor ($C_{\rm em}$) free to vary within a typical range of $0-0.1$ \citep{Mao2022}.
The narrow WE component improves the fitting by $\Delta C / \Delta {\rm DoF} = 28 / 4$, and the best-fit statistic of the final model is $C / {\rm DoF} = 1461/1032$.


\subsection{Stacked High-flux-state Spectrum}

We applied the same continuum model used for the LFS spectrum (i.e., $(dbb+comt+pow)*hot$). The temperature of the Galactic cold gas was fixed at $3.3$~eV according to the best-fit LFS result. The continuum model results in a fitting statistic of $C / {\rm DoF} = 1355/1015$. We do not involve the second {\it hot} component for the Galactic warm-hot halo and the {\it pion} emission component for the warm emitter since the weak features cannot be identified in the low-quality spectrum. A broad Gaussian and two {\it pion} absorption components are still adopted to fit the BEB and WA features of the AGN, with the best-fit LFS parameters to be the initial values. This reduces the fitting statistic to $C / {\rm DoF} = 1242/1004$ ($\Delta C / \Delta {\rm DoF} = 57/4$ for ${\rm WA}_1$, $\Delta C / \Delta {\rm DoF} = 16/4$ for ${\rm WA}_2$, and $\Delta C / \Delta {\rm DoF} = 40/3$ for BEB).

\section{Results and Discussions}
\label{sec:result_discuss}

\begin{figure}
\includegraphics[width=0.47\textwidth]{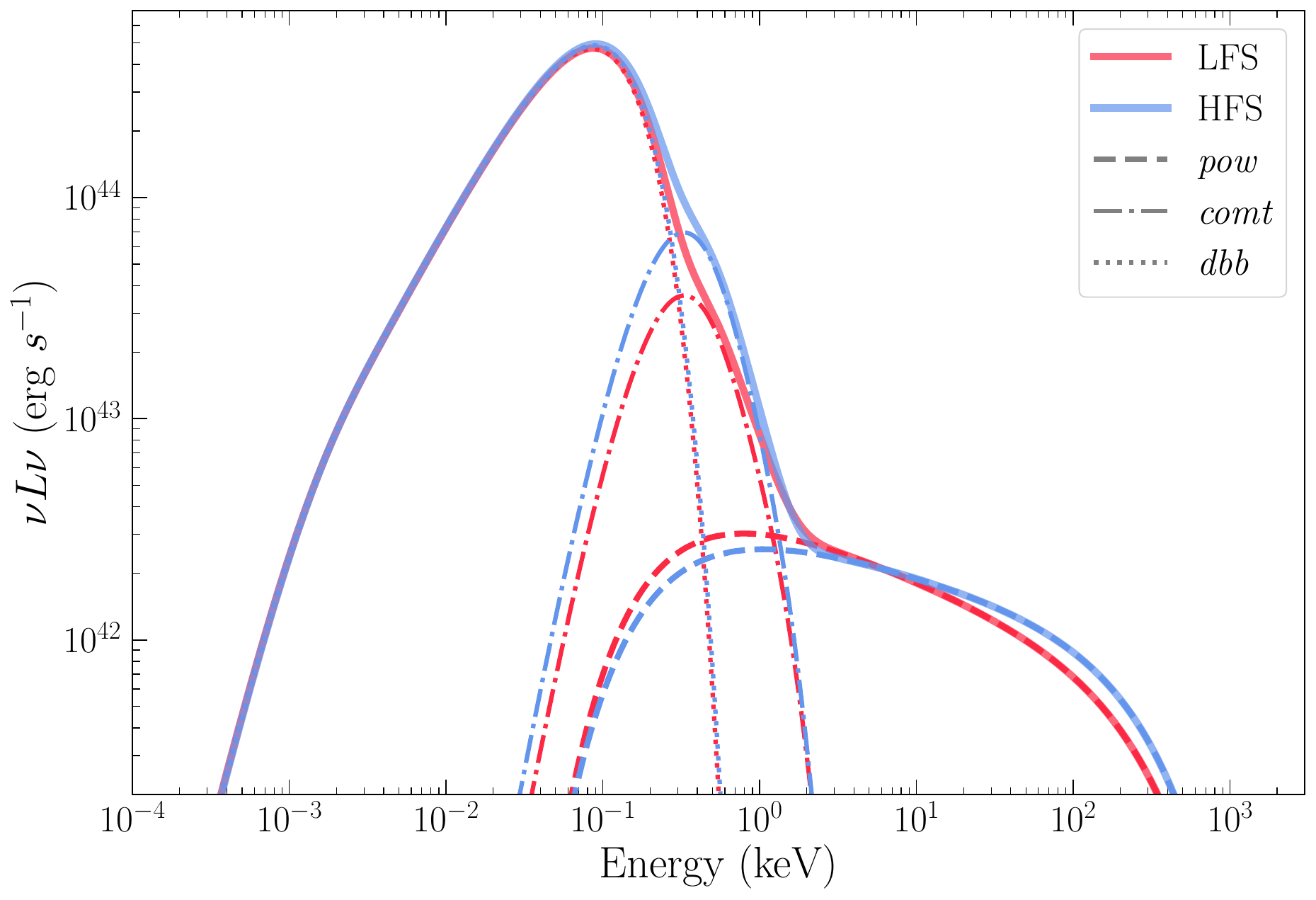}
\caption{Intrinsic SED of \rej. Colors red and blue denote the LFS and HFS, respectively. The solid curves are the best-fit SED, with dashed ({\it pow}), dash-dotted ({\it comt}), and dotted ({\it dbb}) curves its decompositions.}
\label{fig:SED}
\end{figure}

\begin{figure*}
\centering
\includegraphics[width=0.95\textwidth]{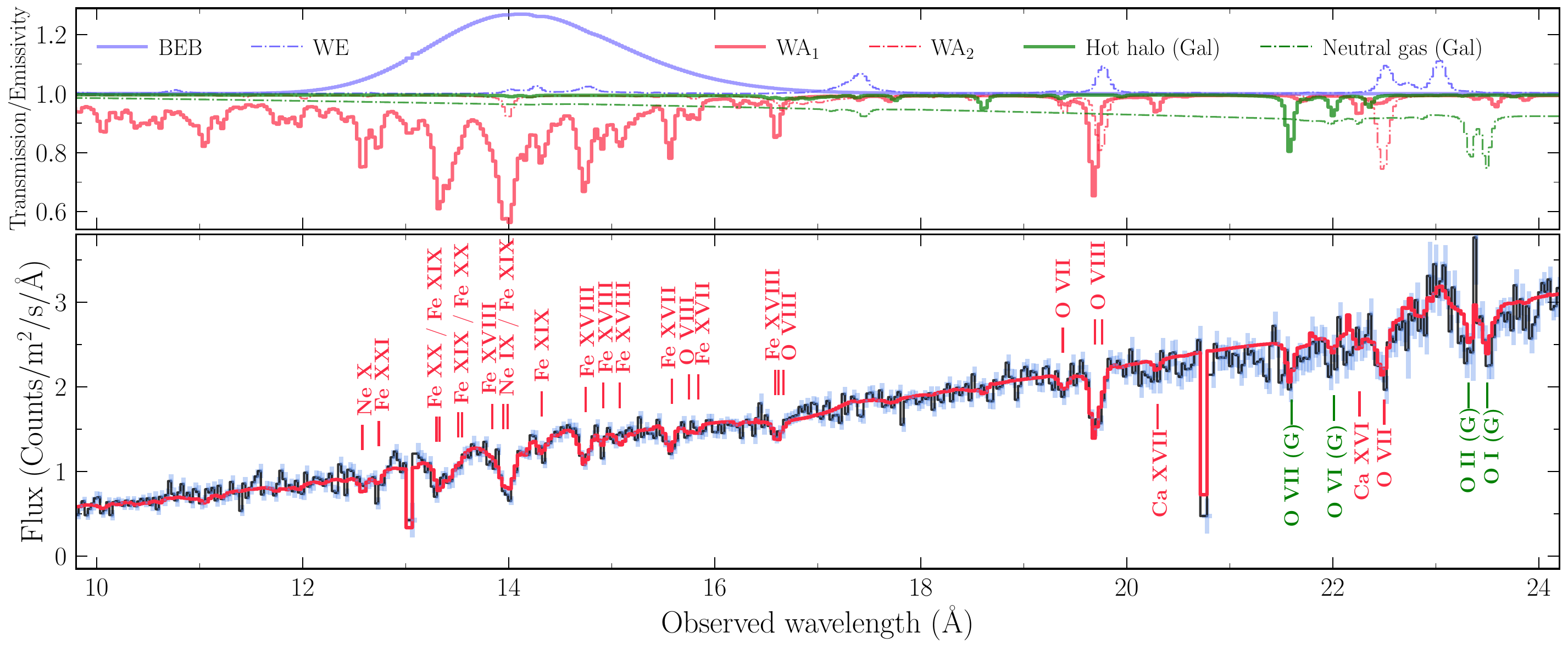} \\
\includegraphics[width=0.95\textwidth]{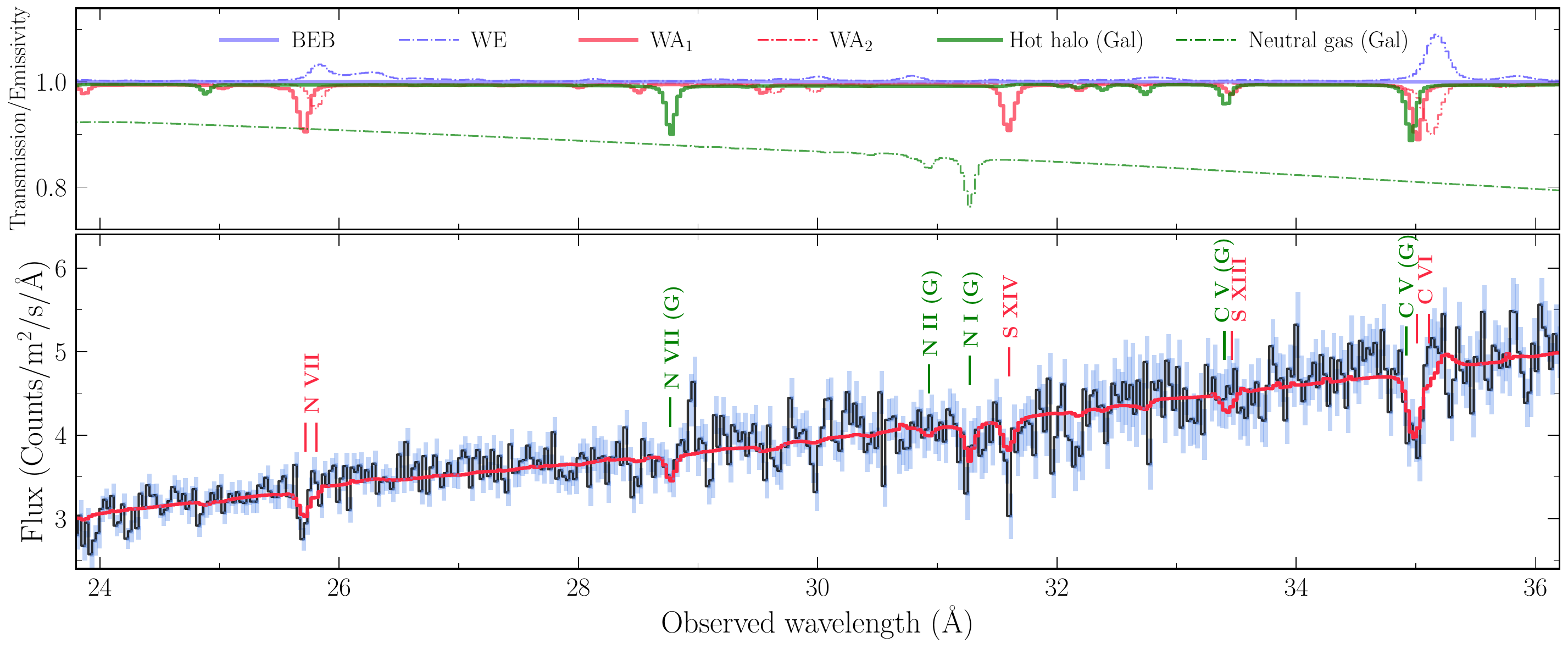} 
\caption{Time-averaged $10-36$~\AA\ LFS RGS spectrum with the best-fit model. The top panel of each segmental spectrum shows the transmission and emissivity of the absorption and emission components, respectively. Colors red and blue label the WA and WE components. Color green marks the Galactic absorption. The bottom panel shows the observed data in black, with $1\,\sigma$ uncertainty in faint blue. All the prominent absorption features are labeled.}
\label{fig:spec_LFS}
\end{figure*}

\begin{figure*}
\centering
\includegraphics[width=0.95\textwidth]{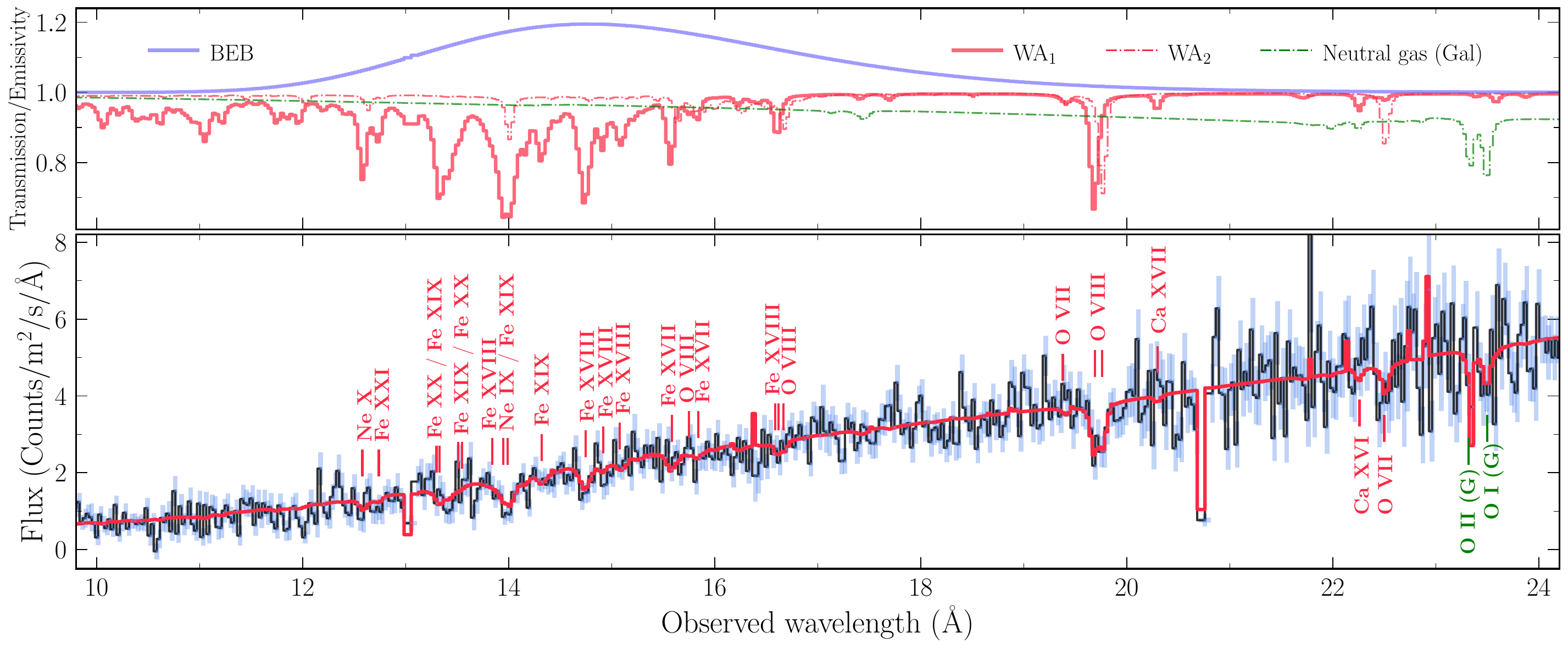} \\
\includegraphics[width=0.95\textwidth]{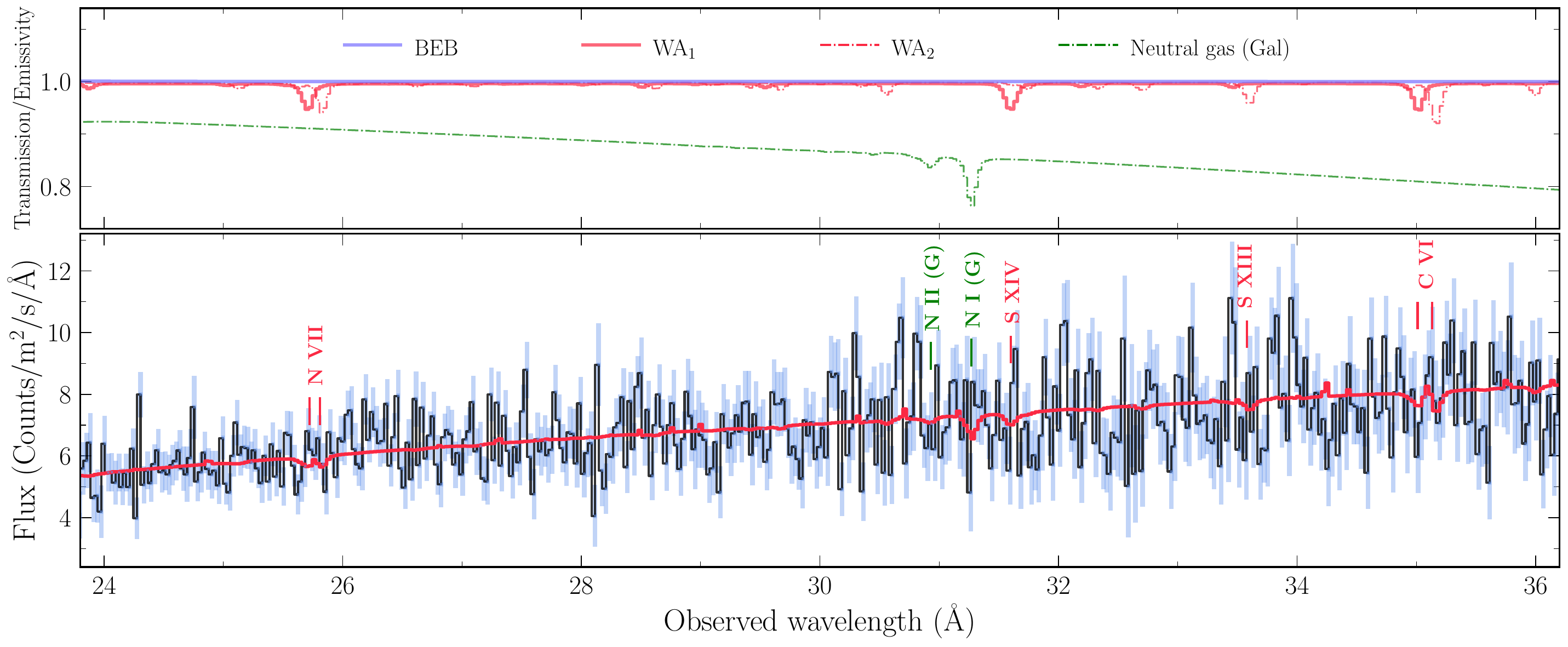} 
\caption{Same as Figure~\ref{fig:spec_LFS}, but for the time-averaged HFS RGS spectrum.}
\label{fig:spec_HFS}
\end{figure*}

We show the optimal model parameters in Table~\ref{tab:result}. Figure~\ref{fig:SED} shows the best-fit SED, and Figures~\ref{fig:spec_LFS} and \ref{fig:spec_HFS} are the best-fit LFS and HFS spectra, respectively.

\renewcommand\arraystretch{1.1}
\begin{deluxetable}{llrr}
\tablewidth{0pt}
\setlength{\tabcolsep}{1 mm}
\tablecaption{Best-fit model parameters. 
\label{tab:result}}
\tablewidth{400pt}
\tablehead{ 
	Comp. & Par. & LFS & HFS  
}
\startdata
\multicolumn{4}{c}{Intrinsic SED} \\
\hline
{\it dbb} & $norm$ ($10^{20}$~m$^2$ ) & $7.31$ (f) & $6.96$ (f) \\
{\it dbb} & $kT_{\rm BB}$ (eV) & $74.5$ (f) & $76.1$ (f)  \\
{\it comt} & $norm$ ($10^{53}$~ph~s$^{-1}$~keV$^{-1}$) & $6.06^{+0.06}_{-0.06}$ & $12.74^{+0.31}_{-0.29}$  \\
{\it comt} & $kT_0$ (eV) & $=kT_{\rm BB}$ & $=kT_{\rm BB}$  \\
{\it comt} & $kT_1$ (keV) & $0.250^{+0.001}_{-0.002}$ & $0.230^{+0.003}_{-0.004}$  \\
{\it comt} & $\tau$ & $11$ (f) & $11$ (f) \\
{\it pow} & $norm$ ($10^{51}$~ph~s$^{-1}$~keV$^{-1}$) & $2.42^{+0.04}_{-0.04}$ & $2.02^{+0.13}_{-0.12}$ \\
{\it pow} & $\Gamma$ & $2.30^{+0.01}_{-0.01}$ & $2.21^{+0.05}_{-0.05}$ \\
\hline
\multicolumn{4}{c}{Galactic neutral gas} \\
\hline
{\it hot} & $\log N_{\rm H}$  (cm$^{-2}$) & $20.10$ (f) & $20.10$ (f)  \\
{\it hot} & $kT$ (eV) & $0.33^{+0.08}_{-0.07}$ & $0.33$ (f)  \\
\hline
\multicolumn{4}{c}{Galactic warm-hot halo} \\
\hline
{\it hot} & $\log N_{\rm H}$  (cm$^{-2}$) & $19.53^{+0.09}_{-0.10}$ & \nodata  \\
{\it hot} & $kT$ (eV) & $30.7^{+3.5}_{-1.9}$ & \nodata  \\
\hline
\multicolumn{4}{c}{Warm absorber \#1} \\
\hline
{\it pion} & $\log N_{\rm H}$ (cm$^{-2}$) & $21.75^{+0.04}_{-0.05}$ & $21.59^{+0.14}_{-0.15}$  \\
{\it pion} & $\log \xi$ (erg~cm~s$^{-1}$) & $3.99^{+0.04}_{-0.04}$ & $3.98^{+0.10}_{-0.11}$ \\
{\it pion} & $v_{\rm out}$ (km~s$^{-1}$) & $-1393^{+44}_{-42}$ & $-1353^{+127}_{-130}$  \\
{\it pion} & $v_{\rm mic}$ (km~s$^{-1}$) & $133^{+31}_{-15}$ & $147^{+126}_{-59}$  \\
{\it pion} & $L_{\rm ion}$ ($10^{44}$~erg~s$^{-1}$)\tablenotemark{a} & $9.65$ & $10.39$  \\
\hline
\multicolumn{4}{c}{Warm absorber \#2} \\
\hline
{\it pion} & $\log N_{\rm H}$ (cm$^{-2}$) & $19.94^{+0.17}_{-0.11}$ & $20.38^{+0.28}_{-0.29}$  \\
{\it pion} & $\log \xi$ (erg~cm~s$^{-1}$) & $2.44^{+0.26}_{-0.15}$ & $2.94^{+0.14}_{-0.14}$   \\
{\it pion} & $v_{\rm out}$ (km~s$^{-1}$) & $-328^{+86}_{-107}$ & $-70^{+170}_{-230}$   \\
{\it pion} & $v_{\rm mic}$ (km~s$^{-1}$) & $425^{+171}_{-134}$ & $133^{+224}_{-74}$  \\
{\it pion}  & $L_{\rm ion}$ ($10^{44}$~erg~s$^{-1}$)\tablenotemark{a} & $9.47$ & $10.25$  \\
\hline
\multicolumn{4}{c}{Broad Emission Bump} \\
\hline
{\it gaus} & $norm$ ($10^{50}$~ph~s$^{-1}$) & $3.41^{+0.28}_{-0.28}$ & $8.32^{+2.11}_{-1.74}$  \\
{\it gaus} & $E$ (keV) & $0.901^{+0.004}_{-0.004}$ & $0.820^{+0.023}_{-0.027}$ \\
{\it gaus} & $FWHM$ (keV) & $0.136^{+0.067}_{-0.010}$ & $0.240^{+0.050}_{-0.039}$ \\
\hline
\multicolumn{4}{c}{Narrow Emission Line} \\
\hline
{\it pion} & $\log N_{\rm H}$ (cm$^{-2}$) & $22.18^{+0.46}_{-0.30}$ & \nodata \\
{\it pion} & $\log \xi$ (erg~cm~s$^{-1}$) & $2.11^{+0.37}_{-0.29}$ & \nodata  \\
{\it pion} & $v_{\rm out}$ (km~s$^{-1}$) & $0$ (f) & \nodata  \\
{\it pion} & $v_{\rm mic}$ (km~s$^{-1}$) & $100$ (f) & \nodata  \\
{\it pion} & $v_{\rm mac}$ (km~s$^{-1}$ ) & $623^{+710}_{-254}$ & \nodata \\
{\it pion} & $C_{\rm em}$ ($\%$) & $3.85^{+2.70}_{-2.36}$ & \nodata \\
\hline
& $C_{\rm stat} / {\rm DoF}$  & $1461/1032$ & $1242/1004$ \\
\enddata
\tablenotetext{a}{$L_{\rm ion}$ is not a fitting parameter but derived from the best-fit model.}
\end{deluxetable}

\subsection{Intrinsic SED}

Between the two flux states, spectral variation focuses on the soft X-ray band and is due to the change of the warm Comptonized disk.
In general, our best-fit SED of both the flux states is consistent with the previous analyses that show a power-law index of $\Gamma \sim 2.2$ and a Compton plasma temperature of $kT_1\sim0.23$~keV \citep{Done2012, Hu2014, Jin2021}. Our best-fit blackbody temperature is $\sim2$ times higher than that in \cite{Jin2021} with a similar model (Model-1 in \citealp{Jin2021}). This discrepancy is due to the difference in the disk blackbody model included in SPEX \citep{Shakura1973} and XSPEC (see, e.g., \citealp{Mitsuda1984}). The bolometric luminosity of the AGN derived by the SED model is $L_{\rm bol}\sim1.2\times10^{45}$~erg~s$^{-1}$ for both the flux states. This value is slightly higher than the previous estimation of $1.04\times10^{45}$~erg~s$^{-1}$ \citep{Jin2012}. Adopting a BH mass of $M_{\rm BH} = 3\times10^6~M_\odot$, the Eddington luminosity of \rej is $L_{\rm Edd} = 3.75 \times10^{44}$~erg~s$^{-1}$ using the empirical equation of $L_{\rm Edd} = 1.25 \times 10^{38} \times (M_{\rm BH} / M_\odot)$ \citep{Rybicki1986}. Thus, the Eddington ratio of \rej is estimated to be $\lambda_{\rm Edd} = L_{\rm bol} / L_{\rm Edd} = 3.2$, with a range of $\lambda_{\rm Edd} \sim 1-10$ due to the mass uncertainty (i.e., $10^6-10^7~M_\odot$). 

\subsection{Absorption from the Milky Way}

We have identified absorption features from the neutral and warm-hot gas components of our Galaxy in the RGS spectra. The best-fit temperature of the neutral component is $0.33^{+0.08}_{-0.07}$~eV ($3.8^{+0.9}_{-0.8}\times10^{3}$~K). This component absorbs the spectral continuum at $>10$~\AA\ areas and generates the absorption lines like O\,{\sc i}, O\,{\sc ii}, N\,{\sc i}, etc. The warm-phase plasma dominates the Galactic warm-hot halo toward the LOS of RE~J1034+396. We find a best-fit temperature of $30.7^{+3.5}_{-1.9}$~eV ($3.6^{+0.4}_{-0.2}\times10^{5}$~K), and an optimal hydrogen column density of $N_{\rm H} = 3.4^{+0.8}_{-0.7}\times10^{19}$~cm$^{-2}$ for the warm-phase halo. As for the hot-phase halo ($10^6-10^7$~K, e.g., \citealp{Fang2015, Nicastro2016, Kaaret2020}), we have not found any significant local features of C\,{\sc vi}, N\,{\sc vii}, and O\,{\sc viii} at a higher temperature. Moreover, the other two useful tracers of Ne\,{\sc ix} and Ne\,{\sc x} (see, e.g., \citealp{Pinto2013, Das2021}) are covered by the WA features, which complicates the decomposition of the hot phase.


\subsection{Warm Absorbers}

\subsubsection{Basic Properties}

We detect two WA components by analyzing the X-ray spectra of RE~J1034+396. The highly ionized component ${\rm WA}_1$ has an ionization parameter of $\log (\xi/{\rm erg~cm~s}^{-1}) \sim 4$ and an outflow velocity of $v_{\rm out} \sim -1400$~km~s$^{-1}$. The main absorption features of ${\rm WA}_1$ are the Fe\,{\sc xvii-xxi} lines at around $14$~\AA\ ($\sim0.9$~keV, observer frame), indicating this is the WA component discovered in previous works \citep{Maitra2010, Middleton2011, Jin2021}. ${\rm WA}_1$ also contributes to the highly ionized lines like O\,{\sc vii-viii}, and Ne\,{\sc ix-x}, etc. The derived $N_{\rm H}$ of ${\rm WA}_1$ in this work is more than two times higher than that in \cite{Middleton2011}. Though we get a consistent $N_{\rm H}$ with \cite{Maitra2010}, the authors fitted the low-resolution EPIC spectrum and interpreted the Fe absorption lines as the O\,{\sc viii} edge. From our high-resolution spectral analysis, our derived ionization parameter is higher than all the previous results with $\log (\xi/{\rm erg~cm~s}^{-1}) \sim 3$ \citep{Maitra2010, Middleton2011}.

${\rm WA}_2$ is a newly discovered component in this work with an outflow velocity of an order $-100$~km~s$^{-1}$. The ionization parameter of this component is $\log (\xi/{\rm erg~cm~s}^{-1}) \sim 2.5-3.0$, and the hydrogen column density is $N_{\rm H}\sim10^{20}$~cm$^{-2}$. ${\rm WA}_2$ contributes to the absorption lines with a moderate ionization state like O\,{\sc vii}, O\,{\sc viii}, C\,{\sc vi}, N\,{\sc vii}, etc.   
Comparing the ${\rm WA}_2$ with ${\rm WA}_1$, we find the WA component with a higher ionization parameter also has higher hydrogen column density and outflow velocity. These correlations have been observed in many other WA targets (e.g., \citealp{Tombesi2013, Laha2014}). The two WA components in \rej manifest a correlation of $v_{\rm out} \propto \xi^{0.4}$ and $v_{\rm out} \propto \xi^{1.2}$ in the LFS and HFS, respectively. The significant change on the index of $v_{\rm out}-\xi$ correlation can not be explained by the pure case of radiatively-driven wind ($v_{\rm out} \propto \xi$, \citealp{King2003}) or magneto-driven wind ($v_{\rm out} \propto \xi^{0.5}$, \citealp{Fukumura2010}). The derived $v_{\rm out}-\xi$ relation is also far from the observed statistical relation of $v \propto \xi^{0.12\pm0.03}$ \citep{Laha2014} and individual cases in \cite{Laha2016} and \cite{Wang2022ApJ}. This result implies a complex launching mechanism for the WAs (see, e.g., \citealp{Laha2016, Wang2022ApJ}). Alternatively, the two WAs in \rej may be separately generated at different times, locations, or with distinct launching mechanisms.

\subsubsection{Radial Location and Mass Outflow Rate}
\label{sec:WA_location}

We estimate the upper and lower distances of the WA winds to the central BH. The upper limit is a geometrical constraint that the thickness of the absorber can not exceed its distance to the central BH (i.e., $\Delta r/r \leq 1$). Combined with the definition of ionization parameter $\xi = L_{\rm ion} / n_{\rm H} r^2$ and $N_{\rm H} = n_{H} \Delta r$, we have $r_{\rm max} = L_{\rm ion} / N_{\rm H} \xi$. The lower boundary is obtained by assuming the outflow velocity is larger than the escape velocity at $r$. Thus, $r_{\rm min} = 2GM_{\rm BH} / v_{\rm out}^2$, where $M_{\rm BH} = 3\times10^6 M_\odot$ is the mass of the central BH, $G$ is the gravitational constant. Substituting the best-fit quantities into the equations, the radial location of ${\rm WA}_1$ is estimated to be $0.013-5.7$~pc by the LFS spectrum and $0.014-9.1$~pc by the HFS spectrum. As for the ${\rm WA}_2$, the corresponding values are $0.24~{\rm pc} - 12.8~{\rm kpc}$ and $5.3~{\rm pc} - 1.6~{\rm kpc}$ constrained by the LFS and HFS data, respectively. Considering the results of both the flux states, the radial location of ${\rm WA}_1$ is estimated to be $0.014~{\rm pc} < r_{\rm WA1} < 5.7~{\rm pc}$ ($\sim 2\times10^4 - 1\times10^7~R_{\rm g}$). The location of ${\rm WA}_2$ is constrained within $5.3~{\rm pc} < r_{\rm WA2} < 1.6~{\rm kpc}$ ($\sim 9\times10^6 - 3\times10^9~R_{\rm g}$).

\begin{figure}
\centering
\includegraphics[width=0.47\textwidth]{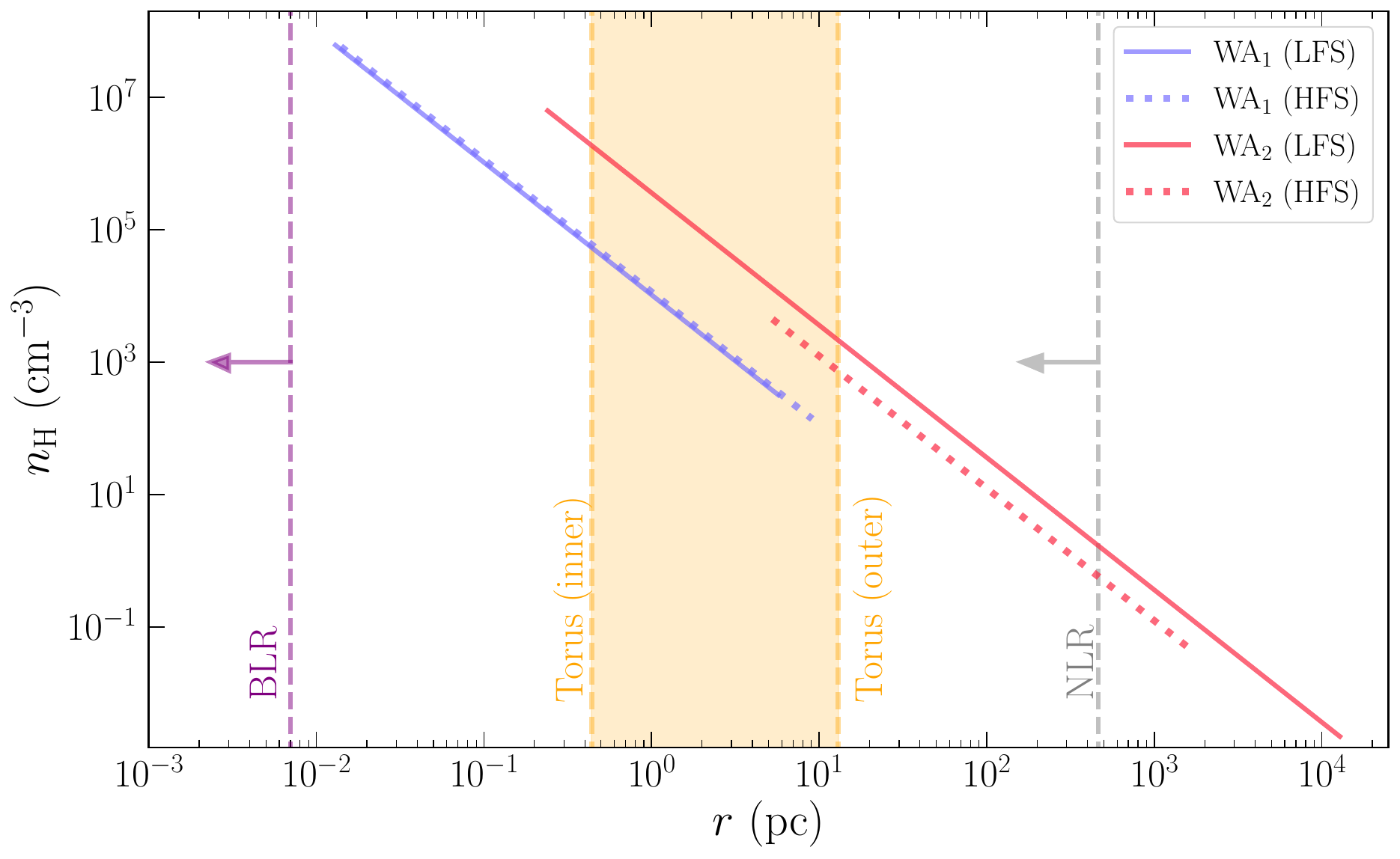}
\caption{Distances of each structure away from the central BH. The blue and red lines show the $n_{\rm H}-R$ solutions of ${\rm WA}_1$ and ${\rm WA}_2$ constrained by the LFS (solid) and HFS (dotted) spectra. The vertical lines mark the locations of BLR, torus, and NLR, respectively.
}
\label{fig:WA_location}
\end{figure}

To illustrate the WA location more intuitively, we roughly calculate the locations of the broad-line region (BLR), dusty torus, and narrow-line region (NLR). The size of BLR is estimated by 
$r_{\rm BLR} = 39.08 \times [\lambda L_\lambda(5100~{\rm \AA}) / (10^{44}\,{\rm erg~s}^{-1}) ]^{0.518}$~light-days \citep{Bentz2006}. Adopting the luminosity derived in \cite{Bian2010} of $\lambda L_\lambda(5100~{\rm \AA}) = 5.6\times10^{42}$~erg~s$^{-1}$, the size of BLR is $0.007$~pc. The inner and outer radii of the torus are calculated according to \cite{Nenkova2008}, where $r_{\rm in} = 0.4 \times (L_{\rm bol} / 10^{45}~{\rm erg~s}^{-1})^{0.5} \times (1500~{\rm K} / T_{\rm d})^{2.6}$~pc, and $r_{\rm out} < 30r_{\rm in}$. Applying a dust sublimation temperature of $T_{\rm d} = 1500$~K, we have $0.44~{\rm pc} < r_{\rm dust} < 13~{\rm pc}$. As for the NLR, its size is estimated by $r_{\rm NLR} = 2.1 ( L_{{\rm [O \ III]}} / 10^{42}~{\rm erg~s}^{-1} )^{0.52} $~kpc \citep{Netzer2004}. An O\,{\sc iii} luminosity of $5.47\times10^{40}$~erg~s$^{-1}$ \citep{Bian2010} leads to a NLR size of $463$~pc. We compare the distances of the WAs to those of BLR, torus, and NLR in Figure~\ref{fig:WA_location}. Both WA components lie outside the BLR. ${\rm WA}_1$ is located within the outer boundary of the torus, while ${\rm WA}_2$ is likely to be located between the inner boundary of the torus and the NLR edge.

The mass outflow rate and the kinetic energy of the WA winds can then be calculated after the location estimation. For a uniform spherical outflow, the mass outflow rate can be estimated by $\dot{M}_{\rm out} = 4 \pi r^2 n_{\rm H} m_{\rm p} v_{\rm out} \leq 4 \pi r m_{\rm p} N_{\rm H} v_{\rm out}$, where $m_{\rm p }$ is the proton mass. Adopting an outflow velocity of $-1400$~km~s$^{-1}$, a hydrogen column density of $10^{21.75}$~cm$^{-2}$, and a maximum location of $5.7$~pc for the ${\rm WA}_1$, the mass outflow rate of ${\rm WA}_1$ is $\dot{M}_{\rm out} < 4.6 ~ M_\odot ~ {\rm yr}^{-1}$. As for the ${\rm WA}_2$, the mass outflow rate is $\dot{M}_{\rm out} < 8.9 ~ M_\odot ~ {\rm yr}^{-1}$ estimated using $v_{\rm out} = -300$~km~s$^{-1}$, $N_{\rm H} = 10^{20.38}$~cm$^{-2}$, and $r_{\rm max} = 1.2$~kpc. The kinetic energy of the WA wind is then $\dot{E}_{\rm K} = 1/2 \dot{M}_{\rm out} v_{\rm out}^2 < 2.9\times10^{42}$~erg~s$^{-1}$ for ${\rm WA}_1$ and $\dot{E}_{\rm K} < 2.4 \times10^{41}$~erg~s$^{-1}$ for ${\rm WA}_2$. Comparing the WA kinetic energy to the bolometric luminosity of RE~J1034+396, we find $\dot{E}_{\rm K} / L_{\rm bol} < 0.24\%$ for ${\rm WA}_1$ and $\dot{E}_{\rm K} / L_{\rm bol} < 0.02\%$ for ${\rm WA}_2$, respectively. The AGN feedback models predict a wind $\dot{E}_{\rm K} / L_{\rm bol}$ ratio of $>0.5\%$ to be an efficient feedback (e.g., \citealp{Hopkins2010}). The low kinetic energy indicates that the WA winds in \rej can not significantly affect the environment of the host galaxy.

\subsection{Time Variation of WA and BEB} 

\subsubsection{WA and BEB in the Two Flux States} 

According to our best-fit models, ${\rm WA}_1$ maintains its properties between the two flux states. In contrast, the ionization parameter of ${\rm WA}_2$ seems to decrease when the source flux state becomes lower. It is against intuition that the outer plasma (i.e., ${\rm WA}_2$) responds to the flux change of the central illuminating source without any variation on the inner plasma (i.e., ${\rm WA}_1$). However, the estimated WA properties of the HFS may be inaccurate due to the low-quality data. A longer exposure of the HFS is required to resolve the WA variation between the different flux states.

The previously discovered WA-QPO connection is based on the analysis of the absorption features at around $14$~\AA\ \citep{Maitra2010, Jin2021}, indicating that it is the ${\rm WA}_1$ potentially exhibits a connection with the QPO phase. It has been reported that the QPO in \rej can only be found at the LFS (e.g., \citealp{Zoghbi2011, Alston2014}). However, our analysis suggests that there might be no difference in the properties of ${\rm WA}_1$ between the two flux states, regardless of the existence of the QPO. This finding casts doubt on the previously discovered connection between the WA and QPO.


We discover a broad emission component centered at $\sim14-15$~\AA\ in both states. The full width at half maxima of BEB is $\sim0.14-0.24$~keV ($\sim45,000-90,000$~km~s$^{-1}$). The BEB shows significant variation in the strength, central wavelength, and broadening between the two states. We attempted to model the BEB in the stacked LFS spectrum using an emitting {\it pion} component (Appendix~\ref{sec:BEB_pion}). The result shows that the BEB could potentially be relativistically broadened features of highly ionized Ne or Fe. However, both explanations require a negligible abundance of oxygen and a very low abundance of metals like C and N. This kind of photoionized emitting plasma has never been discovered in the AGN vicinity. The nature of BEB is still unclear and needs a further check.


The BEB covers the spectral range of $\sim12-18$~\AA\ as shown in Figures~\ref{fig:spec_LFS} and \ref{fig:spec_HFS}. Since the previous works were not aware of the BEB contribution, they made an inaccurate estimation on the continuum level at around $14$~\AA\ for the absorption line analysis. Considering the remarkable variation of BEB in the two flux states, it is likely that ${\rm WA}_1$ does not change its properties in different QPO phases, but the BEB variation leads to the change in the measured equivalent width of the absorption feature. If this is indeed the case, the BEB flux at around $14$~\AA\ should have a positive correlation with the QPO phase. 


\subsubsection{QPO-phase-resolved Spectra of the May-2007 Observation} 


We analyzed the QPO-phase-resolved spectra of the May-2007 observation, in which the WA-QPO connection was first reported. We used the $0.3-10$~keV EPIC-pn light curve to determine the good time intervals of the high QPO phase (HQP) and low QPO phase (LQP) using the method applied in \cite{Maitra2010}. In short, we labeled the flux extremums in each QPO period, and the high QPO intervals were determined when the flux decreases $40\%$ from the peaks to the nearby troughs, while the low QPO intervals were determined when flux increases $40\%$ from the troughs to the nearby peaks. Similar to \cite{Maitra2010}, observational data behind $84$~ks are excluded due to soft-proton flares. 

\begin{figure*}
\centering
\includegraphics[width=0.47\textwidth]{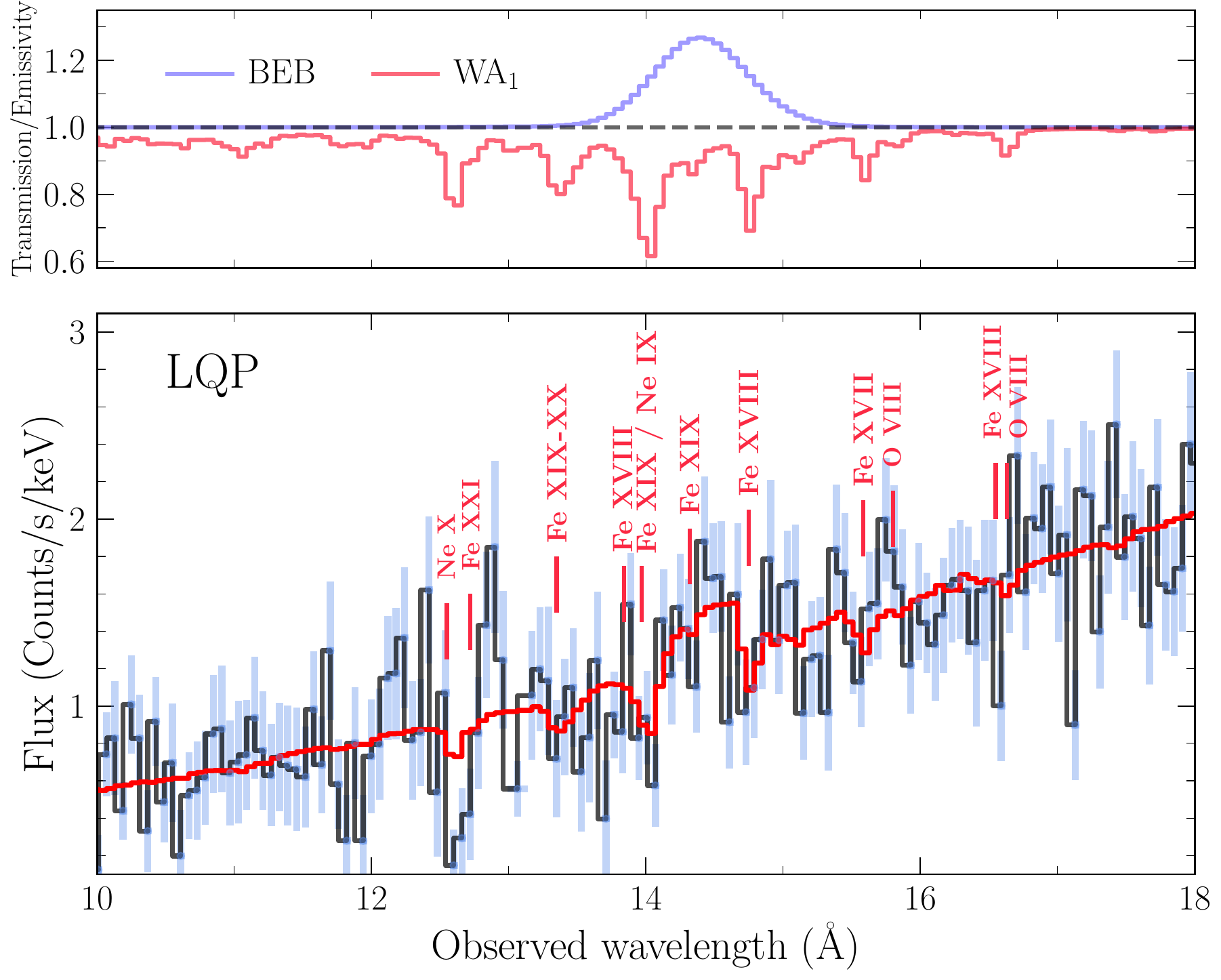}
\includegraphics[width=0.47\textwidth]{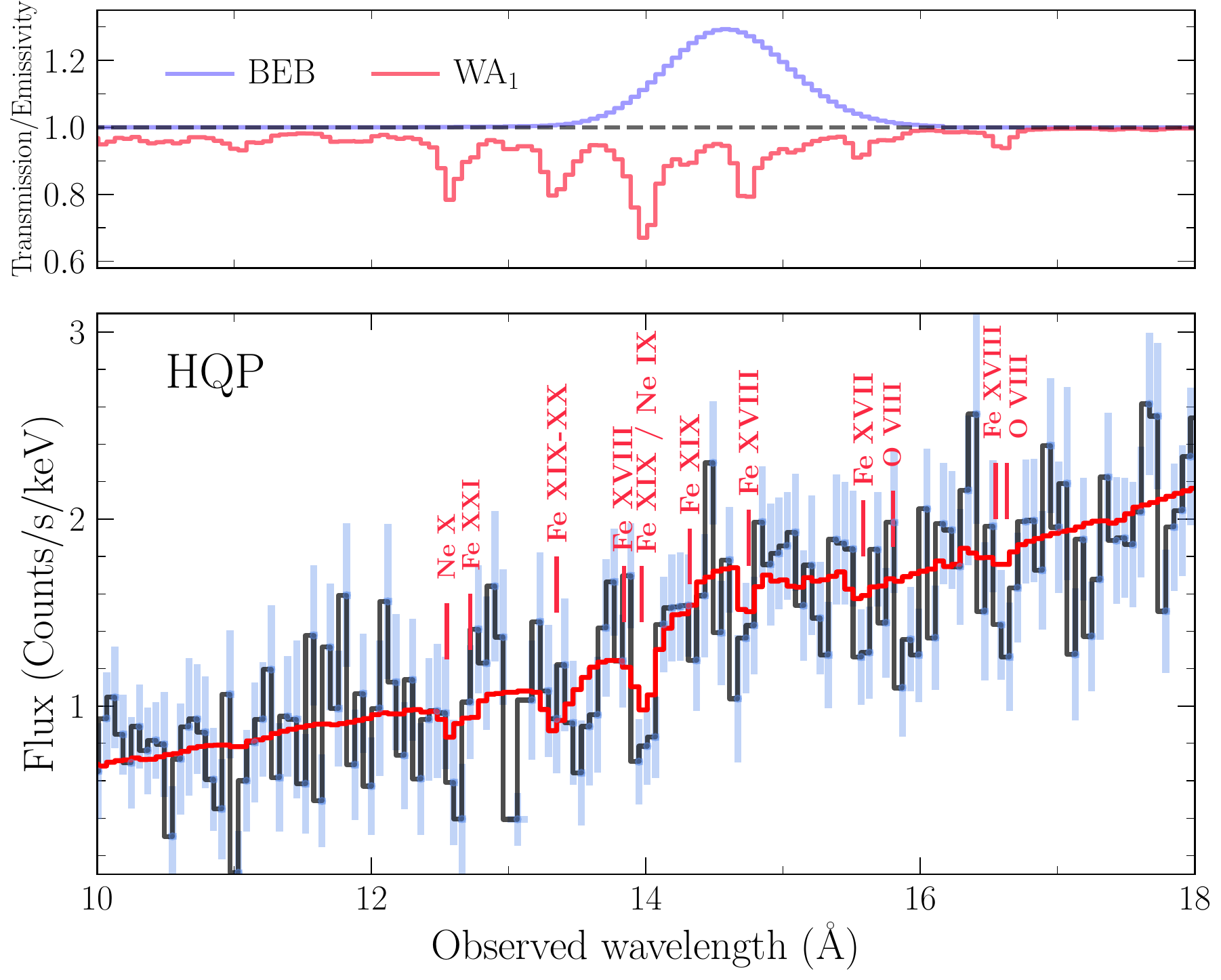}
\caption{$10-18$~\AA\ RGS spectra of the LQP (left) and HQP (right) of the May-2007 observation. The top panel shows the model decompositions, with the BEB in blue and the high-ionization WA in red. The bottom panel shows the observed data in black and the $1\sigma$ uncertainty in blue. The red curve shows the best-fit model.}
\label{fig:phase_spec}
\end{figure*}

We extracted the QPO-phase-resolved spectra of RGS and EPIC-pn using the derived phase intervals. The net exposures of the HQP and LQP are $32.6$~ks and $29.3$~ks, respectively. We re-binned the RGS spectra by a factor of six due to the poor data quality. The RGS and EPIC-pn spectra combined with the $UVW1$ count rates were jointly analyzed. The spectral continuum was fitted by {\it (pow+comt+dbb)*hot}, with the Galactic neutral gas absorption fixed at the best-fit stacked LFS value. Absorption from ${\rm WA}_1$ and emission from BEB were fitted by a {\it pion} component and a Gaussian profile, respectively. We did not model the features from Galactic warm-hot gas, ${\rm WA}_2$, and warm emitter due to their negligible contribution to the low-quality spectra ($\Delta C\sim0$). The best-fit model parameters and $10-18$~\AA\ RGS spectrum are shown in Table~\ref{tab:result_phase_spec} and Figure~\ref{fig:phase_spec}, respectively.

\renewcommand\arraystretch{1.1}
\begin{deluxetable}{llrr}
\tablewidth{0pt}
\setlength{\tabcolsep}{1 mm}
\tablecaption{Best-fit models of the QPO-phase-resolved spectra. 
\label{tab:result_phase_spec}}
\tablewidth{400pt}
\tablehead{ 
	Comp. & Par. & LQP & HQP  
}
\startdata
\multicolumn{4}{c}{Intrinsic SED} \\
\hline
{\it dbb} & $norm$ ($10^{20}$~m$^2$ ) & $5.88$ (f) & $6.22$ (f) \\
{\it dbb} & $kT_{\rm BB}$ (eV) & $80.4$ (f) & $78.8$ (f)  \\
{\it comt} & $norm$ ($10^{53}$~ph~s$^{-1}$~keV$^{-1}$) & $5.86^{+0.23}_{-0.22}$ & $6.99^{+0.32}_{-0.30}$  \\
{\it comt} & $kT_0$ (eV) & $=kT_{\rm BB}$ & $=kT_{\rm BB}$  \\
{\it comt} & $kT_1$ (keV) & $0.247^{+0.005}_{-0.005}$ & $0.228^{+0.006}_{-0.006}$  \\
{\it comt} & $\tau$ & $11$ (f) & $11$ (f) \\
{\it pow} & $norm$ ($10^{51}$~ph~s$^{-1}$~keV$^{-1}$) & $1.61^{+0.14}_{-0.13}$ & $3.35^{+0.19}_{-0.18}$ \\
{\it pow} & $\Gamma$ & $2.16^{+0.07}_{-0.07}$ & $2.42^{+0.05}_{-0.05}$ \\
\hline
\multicolumn{4}{c}{Warm absorber \#1} \\
\hline
{\it pion} & $\log N_{\rm H}$ (cm$^{-2}$) & $21.39^{+0.16}_{-0.48}$ & $21.35^{+0.17}_{-0.22}$  \\
{\it pion} & $\log \xi$ (erg~cm~s$^{-1}$) & $4.03^{+0.18}_{-0.14}$ & $3.95^{+0.13}_{-0.15}$ \\
{\it pion} & $v_{\rm out}$ (km~s$^{-1}$) & $-575^{+327}_{-334}$ & $-1236^{+439}_{-471}$  \\
{\it pion} & $v_{\rm mic}$ (km~s$^{-1}$) & $529^{+426}_{-287}$ & $765^{+520}_{-646}$  \\
\hline
\multicolumn{4}{c}{Broad Emission Bump} \\
\hline
{\it gaus} & $norm$ ($10^{50}$~ph~s$^{-1}$) & $1.31^{+0.73}_{-0.98}$ & $2.10^{+0.83}_{-0.66}$  \\
{\it gaus} & $E$ (keV) & $0.895^{+0.012}_{-0.013}$ & $0.882^{+0.011}_{-0.009}$ \\
{\it gaus} & $FWHM$ (keV) & $0.046^{+0.029}_{-0.046}$ & $0.061^{+0.024}_{-0.019}$ \\
\hline
& $C_{\rm stat} / {\rm DoF}$  & $589/540$ & $588/541$ \\
\enddata
\end{deluxetable}

The two-phase spectra exhibit a significant difference in the power-law component, with the HQP showing a steeper spectrum and a larger power-law normalization. This finding is consistent with previous timing analysis, which suggested the power-law component is likely to take responsibility for the periodic flux change \citep{Middleton2009, Zoghbi2011}. We detect the ${\rm WA}_1$ in both QPO phases without any obvious difference, in contrast to the previous results showing a non-detection of ${\rm WA}_1$ in the HQP spectrum \citep{Maitra2010, Jin2021}. The BEB seems stronger (in normalization) and broader in the HQP than the LQP, implying a positive correlation between BEB intensity and the QPO phase. Therefore, the dramatic variation of BEB and the unawareness of BEB contribution in the early works may lead to the wrong impression of the WA-QPO connection. The correlation between BEB and QPO is not assertive due to poor data quality. A detailed analysis of the BEB nature and its timing properties will be presented in our future work on the target.



\section{Summary}

In this work, we conduct a detailed analysis of WA winds in RE~J1034+396 with more than $1$~Ms \xmm observations. We analyzed the properties of WA winds in low and high flux states as well as the May-2007 observation in low and high QPO phases. Our main findings are summarized as follows.

\begin{enumerate}

\item Two WA components are required to explain the intrinsic absorption features in the time-averaged RGS spectra. The highly ionized component, which has been discovered before \citep{Maitra2010,Middleton2011,Jin2021}, has an ionization parameter of $\log (\xi/{\rm erg~cm~s}^{-1}) \sim 4$, an outflow velocity of around $-1400$~km~s$^{-1}$, and contributes to the transitions like O\,{\sc vii-viii}, Ne\,{\sc ix-x}, and Fe\,{\sc xvii-xxi}. The lower ionized component is newly discovered in this work. This component shows an outflow velocity of around $-(100-300)$~km~s$^{-1}$ and a ionization parameter of $\log (\xi/{\rm erg~cm~s}^{-1}) \sim 2.5-3$. It primarily models the absorption features of O\,{\sc vii-viii}, N\,{\sc vii}, and C\,{\sc vi}.

\item The highly ionized WA is likely to be located within the outer boundary of the torus but outside the BLR. While the low-ionization WA may be located between the inner boundary of the torus and the NLR. 
The kinetic energy of the wind is estimated to be $<0.24\% L_{\rm bol}$ and $<0.02\% L_{\rm bol}$ for the high- and low-ionization WAs, respectively, which is unlikely to produce a significant impact on the host galaxy.

\item We find no difference in the highly ionized WA between the two flux states and the two QPO phases, which is against the WA-QPO connection discovered in previous works. The ionization parameter of the low-ionization WA may positively correlate with the source flux. The suspected correlation may be biased by the poor data quality of the HFS and should be further explored with deeper observations.

\item We identify a broad emission bump at around $14$~\AA, which covers the primary features of the high-ionization WA. This component shows a significant variation between the two flux states, and its strength may positively correlate with the QPO phase. The dramatic variation of the broad emission bump may take responsibility for the misidentified WA-QPO connection in the early works.

\item The cold and warm-hot halos of the MW contribute to the absorption features seen in the RGS spectra. The best-fit temperatures of the two components are $\sim3.8\times10^3$~K and $\sim3.6\times10^5$~K, respectively.

\end{enumerate}

We expect that our study will benefit future analyses on both the absorption and emission phenomena of the target. By conducting QPO-phase-resolved spectroscopy with more than $1$~Ms \xmm observations, future studies may provide a better understanding of the BEB nature and its potential correlation with the QPO, which will empower us to investigate the vicinity of the supermassive black hole with deeper insight.

\begin{acknowledgments}
We thank the anonymous referee for careful reading and helpful comments that improved the paper. The data and scripts used in this work are available at Zenodo DOI: \href{https://doi.org/10.5281/zenodo.10866089}{10.5281/zenodo.10866089.} 
Z.Z appreciates Dr. Qingzheng Yu and Honghui Liu for their helpful suggestions. This work is supported by the National Key R\&D Program of China under No. 2017YFA0402600, and the National Natural Science Foundation of China under Nos. 11890692, 12133008, 12221003. We acknowledge the science research grant from the China Manned Space Project with No. CMS-CSST-2021-A04.
\end{acknowledgments}

%


\bibliography{ref}{}

\begin{thebibliography}{}
\expandafter\ifx\csname natexlab\endcsname\relax\def\natexlab#1{#1}\fi
\providecommand{\url}[1]{\href{#1}{#1}}
\providecommand{\dodoi}[1]{doi:~\href{http://doi.org/#1}{\nolinkurl{#1}}}
\providecommand{\doeprint}[1]{\href{http://ascl.net/#1}{\nolinkurl{http://ascl.net/#1}}}
\providecommand{\doarXiv}[1]{\href{https://arxiv.org/abs/#1}{\nolinkurl{https://arxiv.org/abs/#1}}}

\bibitem[{{Alston} {et~al.}(2014){Alston}, {Markeviciute}, {Kara}, {Fabian}, \&
  {Middleton}}]{Alston2014}
{Alston}, W.~N., {Markeviciute}, J., {Kara}, E., {Fabian}, A.~C., \&
  {Middleton}, M. 2014, \mnras, 445, L16, \dodoi{10.1093/mnrasl/slu127}

\bibitem[{{Behar} {et~al.}(2001){Behar}, {Sako}, \& {Kahn}}]{Behar2001}
{Behar}, E., {Sako}, M., \& {Kahn}, S.~M. 2001, \apj, 563, 497,
  \dodoi{10.1086/323966}

\bibitem[{{Bentz} {et~al.}(2006){Bentz}, {Peterson}, {Pogge}, {Vestergaard}, \&
  {Onken}}]{Bentz2006}
{Bentz}, M.~C., {Peterson}, B.~M., {Pogge}, R.~W., {Vestergaard}, M., \&
  {Onken}, C.~A. 2006, \apj, 644, 133, \dodoi{10.1086/503537}

\bibitem[{{Bian} \& {Huang}(2010)}]{Bian2010}
{Bian}, W.-H., \& {Huang}, K. 2010, \mnras, 401, 507,
  \dodoi{10.1111/j.1365-2966.2009.15662.x}

\bibitem[{{Blandford} \& {Payne}(1982)}]{Blandford1982}
{Blandford}, R.~D., \& {Payne}, D.~G. 1982, \mnras, 199, 883,
  \dodoi{10.1093/mnras/199.4.883}

\bibitem[{{Blustin} {et~al.}(2005){Blustin}, {Page}, {Fuerst},
  {Branduardi-Raymont}, \& {Ashton}}]{Blustin2005}
{Blustin}, A.~J., {Page}, M.~J., {Fuerst}, S.~V., {Branduardi-Raymont}, G., \&
  {Ashton}, C.~E. 2005, \aap, 431, 111, \dodoi{10.1051/0004-6361:20041775}

\bibitem[{{Bottorff} {et~al.}(2000){Bottorff}, {Korista}, \&
  {Shlosman}}]{Bottorff2000}
{Bottorff}, M.~C., {Korista}, K.~T., \& {Shlosman}, I. 2000, \apj, 537, 134,
  \dodoi{10.1086/309006}

\bibitem[{{Buhariwalla} {et~al.}(2020){Buhariwalla}, {Waddell}, {Gallo},
  {Grupe}, \& {Komossa}}]{Buhariwalla2020}
{Buhariwalla}, M.~Z., {Waddell}, S. G.~H., {Gallo}, L.~C., {Grupe}, D., \&
  {Komossa}, S. 2020, \apj, 901, 118, \dodoi{10.3847/1538-4357/abb08a}

\bibitem[{{Burnham} {et~al.}(2011){Burnham}, {Anderson}, \&
  {Huyvaert}}]{Burnham2011}
{Burnham}, K.~P., {Anderson}, D.~R., \& {Huyvaert}, K.~P. 2011, {\it Behav Ecol
  Sociobiol}, 65, 23, \dodoi{https://doi.org/10.1007/s00265-010-1029-6}

\bibitem[{{Chaudhury} {et~al.}(2018){Chaudhury}, {Chitnis}, {Rao}, {Singh},
  {Bhattacharyya}, {Dewangan}, {Chakraborty}, {Chandra}, {Stewart}, {Mukerjee},
  \& {Dey}}]{Chaudhury2018}
{Chaudhury}, K., {Chitnis}, V.~R., {Rao}, A.~R., {et~al.} 2018, \mnras, 478,
  4830, \dodoi{10.1093/mnras/sty1366}

\bibitem[{{Czerny} {et~al.}(2016){Czerny}, {You}, {Kurcz},
  {{\'S}redzi{\'n}ska}, {Hryniewicz}, {Niko{\l}ajuk}, {Krupa}, {Wang}, {Hu}, \&
  {{\.Z}ycki}}]{Czerny2016}
{Czerny}, B., {You}, B., {Kurcz}, A., {et~al.} 2016, \aap, 594, A102,
  \dodoi{10.1051/0004-6361/201628103}

\bibitem[{{Dannen} {et~al.}(2019){Dannen}, {Proga}, {Kallman}, \&
  {Waters}}]{Dannen2019}
{Dannen}, R.~C., {Proga}, D., {Kallman}, T.~R., \& {Waters}, T. 2019, \apj,
  882, 99, \dodoi{10.3847/1538-4357/ab340b}

\bibitem[{{Das} {et~al.}(2021){Das}, {Mathur}, {Gupta}, \&
  {Krongold}}]{Das2021}
{Das}, S., {Mathur}, S., {Gupta}, A., \& {Krongold}, Y. 2021, \apj, 918, 83,
  \dodoi{10.3847/1538-4357/ac0e8e}

\bibitem[{{de Plaa} {et~al.}(2004){de Plaa}, {Kaastra}, {Tamura},
  {Pointecouteau}, {Mendez}, \& {Peterson}}]{de_Plaa2004}
{de Plaa}, J., {Kaastra}, J.~S., {Tamura}, T., {et~al.} 2004, \aap, 423, 49,
  \dodoi{10.1051/0004-6361:20047170}

\bibitem[{{Done} {et~al.}(2012){Done}, {Davis}, {Jin}, {Blaes}, \&
  {Ward}}]{Done2012}
{Done}, C., {Davis}, S.~W., {Jin}, C., {Blaes}, O., \& {Ward}, M. 2012, \mnras,
  420, 1848, \dodoi{10.1111/j.1365-2966.2011.19779.x}

\bibitem[{{Dorodnitsyn} {et~al.}(2008){Dorodnitsyn}, {Kallman}, \&
  {Proga}}]{Dorodnitsyn2008}
{Dorodnitsyn}, A., {Kallman}, T., \& {Proga}, D. 2008, \apj, 687, 97,
  \dodoi{10.1086/591418}

\bibitem[{{Ebrero} {et~al.}(2011){Ebrero}, {Kriss}, {Kaastra}, {Detmers},
  {Steenbrugge}, {Costantini}, {Arav}, {Bianchi}, {Cappi},
  {Branduardi-Raymont}, {Mehdipour}, {Petrucci}, {Pinto}, \&
  {Ponti}}]{Ebrero2011}
{Ebrero}, J., {Kriss}, G.~A., {Kaastra}, J.~S., {et~al.} 2011, \aap, 534, A40,
  \dodoi{10.1051/0004-6361/201117067}

\bibitem[{{Fang} {et~al.}(2015){Fang}, {Buote}, {Bullock}, \& {Ma}}]{Fang2015}
{Fang}, T., {Buote}, D., {Bullock}, J., \& {Ma}, R. 2015, \apjs, 217, 21,
  \dodoi{10.1088/0067-0049/217/2/21}

\bibitem[{{Fukumura} {et~al.}(2010){Fukumura}, {Kazanas}, {Contopoulos}, \&
  {Behar}}]{Fukumura2010}
{Fukumura}, K., {Kazanas}, D., {Contopoulos}, I., \& {Behar}, E. 2010, \apj,
  715, 636, \dodoi{10.1088/0004-637X/715/1/636}

\bibitem[{{Fukumura} {et~al.}(2018){Fukumura}, {Kazanas}, {Shrader}, {Behar},
  {Tombesi}, \& {Contopoulos}}]{Fukumura2018}
{Fukumura}, K., {Kazanas}, D., {Shrader}, C., {et~al.} 2018, \apj, 853, 40,
  \dodoi{10.3847/1538-4357/aaa3f6}

\bibitem[{{Gierli{\'n}ski} {et~al.}(2008){Gierli{\'n}ski}, {Middleton}, {Ward},
  \& {Done}}]{Gierliski2008}
{Gierli{\'n}ski}, M., {Middleton}, M., {Ward}, M., \& {Done}, C. 2008, \nat,
  455, 369, \dodoi{10.1038/nature07277}

\bibitem[{{Gnat} \& {Sternberg}(2007)}]{Gnat2007}
{Gnat}, O., \& {Sternberg}, A. 2007, \apjs, 168, 213, \dodoi{10.1086/509786}

\bibitem[{{Gonzalez} {et~al.}(2018){Gonzalez}, {Waddell}, \&
  {Gallo}}]{Gonzalez2018}
{Gonzalez}, A.~G., {Waddell}, S.~G.~H., \& {Gallo}, L.~C. 2018, \mnras, 475,
  128, \dodoi{10.1093/mnras/stx3146}

\bibitem[{{Gupta} {et~al.}(2013){Gupta}, {Mathur}, {Krongold}, \&
  {Nicastro}}]{Gupta2013}
{Gupta}, A., {Mathur}, S., {Krongold}, Y., \& {Nicastro}, F. 2013, \apj, 768,
  141, \dodoi{10.1088/0004-637X/768/2/141}

\bibitem[{{Halpern}(1984)}]{Halpern1984}
{Halpern}, J.~P. 1984, \apj, 281, 90, \dodoi{10.1086/162077}

\bibitem[{{HI4PI Collaboration} {et~al.}(2016){HI4PI Collaboration}, {Ben
  Bekhti}, {Fl{\"o}er}, {Keller}, {Kerp}, {Lenz}, {Winkel}, {Bailin},
  {Calabretta}, {Dedes}, {Ford}, {Gibson}, {Haud}, {Janowiecki}, {Kalberla},
  {Lockman}, {McClure-Griffiths}, {Murphy}, {Nakanishi}, {Pisano}, \&
  {Staveley-Smith}}]{HI4PI}
{HI4PI Collaboration}, {Ben Bekhti}, N., {Fl{\"o}er}, L., {et~al.} 2016, \aap,
  594, A116, \dodoi{10.1051/0004-6361/201629178}

\bibitem[{{Hopkins} \& {Elvis}(2010)}]{Hopkins2010}
{Hopkins}, P.~F., \& {Elvis}, M. 2010, \mnras, 401, 7,
  \dodoi{10.1111/j.1365-2966.2009.15643.x}

\bibitem[{{Hu} {et~al.}(2014){Hu}, {Chou}, {Yang}, \& {Su}}]{Hu2014}
{Hu}, C.-P., {Chou}, Y., {Yang}, T.-C., \& {Su}, Y.-H. 2014, \apj, 788, 31,
  \dodoi{10.1088/0004-637X/788/1/31}

\bibitem[{{Huppenkothen} {et~al.}(2019){Huppenkothen}, {Bachetti}, {Stevens},
  {Migliari}, {Balm}, {Hammad}, {Khan}, {Mishra}, {Rashid}, {Sharma}, {Martinez
  Ribeiro}, \& {Valles Blanco}}]{Huppenkothen2019}
{Huppenkothen}, D., {Bachetti}, M., {Stevens}, A.~L., {et~al.} 2019, \apj, 881,
  39, \dodoi{10.3847/1538-4357/ab258d}

\bibitem[{{Jin} {et~al.}(2020){Jin}, {Done}, \& {Ward}}]{Jin2020}
{Jin}, C., {Done}, C., \& {Ward}, M. 2020, \mnras, 495, 3538,
  \dodoi{10.1093/mnras/staa1356}

\bibitem[{{Jin} {et~al.}(2021){Jin}, {Done}, \& {Ward}}]{Jin2021}
---. 2021, \mnras, 500, 2475, \dodoi{10.1093/mnras/staa3386}

\bibitem[{{Jin} {et~al.}(2012){Jin}, {Ward}, {Done}, \& {Gelbord}}]{Jin2012}
{Jin}, C., {Ward}, M., {Done}, C., \& {Gelbord}, J. 2012, \mnras, 420, 1825,
  \dodoi{10.1111/j.1365-2966.2011.19805.x}

\bibitem[{{Kaaret} {et~al.}(2020){Kaaret}, {Koutroumpa}, {Kuntz}, {Jahoda},
  {Bluem}, {Gulick}, {Hodges-Kluck}, {LaRocca}, {Ringuette}, \&
  {Zajczyk}}]{Kaaret2020}
{Kaaret}, P., {Koutroumpa}, D., {Kuntz}, K.~D., {et~al.} 2020, Nature
  Astronomy, 4, 1072, \dodoi{10.1038/s41550-020-01215-w}

\bibitem[{{Kaastra}(2017)}]{Kaastra2017}
{Kaastra}, J.~S. 2017, \aap, 605, A51, \dodoi{10.1051/0004-6361/201629319}

\bibitem[{{Kaastra} \& {Bleeker}(2016)}]{Kaastra2016}
{Kaastra}, J.~S., \& {Bleeker}, J.~A.~M. 2016, \aap, 587, A151,
  \dodoi{10.1051/0004-6361/201527395}

\bibitem[{{Kaastra} {et~al.}(2000){Kaastra}, {Mewe}, {Liedahl}, {Komossa}, \&
  {Brinkman}}]{Kaastra2000}
{Kaastra}, J.~S., {Mewe}, R., {Liedahl}, D.~A., {Komossa}, S., \& {Brinkman},
  A.~C. 2000, \aap, 354, L83, \dodoi{10.48550/arXiv.astro-ph/0002345}

\bibitem[{{Kaastra} {et~al.}(1996){Kaastra}, {Mewe}, \&
  {Nieuwenhuijzen}}]{Kaastra1996}
{Kaastra}, J.~S., {Mewe}, R., \& {Nieuwenhuijzen}, H. 1996, in UV and X-ray
  Spectroscopy of Astrophysical and Laboratory Plasmas, 411--414

\bibitem[{{Kaastra} {et~al.}(2020){Kaastra}, {Raassen}, {de Plaa}, \&
  {Gu}}]{Kaastra2020}
{Kaastra}, J.~S., {Raassen}, A.~J.~J., {de Plaa}, J., \& {Gu}, L. 2020, {SPEX
  X-ray spectral fitting package}, 3.06.01, Zenodo,  Zenodo,
  \dodoi{10.5281/zenodo.4384188}

\bibitem[{{Kaastra} {et~al.}(2004){Kaastra}, {Raassen}, {Mewe}, {Arav},
  {Behar}, {Costantini}, {Gabel}, {Kriss}, {Proga}, {Sako}, \&
  {Steenbrugge}}]{Kaastra2004}
{Kaastra}, J.~S., {Raassen}, A.~J.~J., {Mewe}, R., {et~al.} 2004, \aap, 428,
  57, \dodoi{10.1051/0004-6361:20041434}

\bibitem[{{Kaastra} {et~al.}(2012){Kaastra}, {Detmers}, {Mehdipour}, {Arav},
  {Behar}, {Bianchi}, {Branduardi-Raymont}, {Cappi}, {Costantini}, {Ebrero},
  {Kriss}, {Paltani}, {Petrucci}, {Pinto}, {Ponti}, {Steenbrugge}, \& {de
  Vries}}]{Kaastra2012}
{Kaastra}, J.~S., {Detmers}, R.~G., {Mehdipour}, M., {et~al.} 2012, \aap, 539,
  A117, \dodoi{10.1051/0004-6361/201118161}

\bibitem[{{Kaspi} {et~al.}(2002){Kaspi}, {Brandt}, {George}, {Netzer},
  {Crenshaw}, {Gabel}, {Hamann}, {Kaiser}, {Koratkar}, {Kraemer}, {Kriss},
  {Mathur}, {Mushotzky}, {Nandra}, {Peterson}, {Shields}, {Turner}, \&
  {Zheng}}]{Kaspi2002}
{Kaspi}, S., {Brandt}, W.~N., {George}, I.~M., {et~al.} 2002, \apj, 574, 643,
  \dodoi{10.1086/341113}

\bibitem[{{Kaufman} {et~al.}(2017){Kaufman}, {Blaes}, \&
  {Hirose}}]{Kaufman2017}
{Kaufman}, J., {Blaes}, O.~M., \& {Hirose}, S. 2017, \mnras, 467, 1734,
  \dodoi{10.1093/mnras/stx193}

\bibitem[{{Khanna} {et~al.}(2016){Khanna}, {Kaastra}, \&
  {Mehdipour}}]{Khanna2016}
{Khanna}, S., {Kaastra}, J.~S., \& {Mehdipour}, M. 2016, \aap, 586, A2,
  \dodoi{10.1051/0004-6361/201527047}

\bibitem[{{King} \& {Pounds}(2015)}]{King2015}
{King}, A., \& {Pounds}, K. 2015, \araa, 53, 115,
  \dodoi{10.1146/annurev-astro-082214-122316}

\bibitem[{{King} \& {Pounds}(2003)}]{King2003}
{King}, A.~R., \& {Pounds}, K.~A. 2003, \mnras, 345, 657,
  \dodoi{10.1046/j.1365-8711.2003.06980.x}

\bibitem[{{Kormendy} \& {Ho}(2013)}]{Kormendy2013}
{Kormendy}, J., \& {Ho}, L.~C. 2013, \araa, 51, 511,
  \dodoi{10.1146/annurev-astro-082708-101811}

\bibitem[{{Krolik} \& {Kriss}(2001)}]{Krolik2001}
{Krolik}, J.~H., \& {Kriss}, G.~A. 2001, \apj, 561, 684, \dodoi{10.1086/323442}

\bibitem[{{Krongold} {et~al.}(2003){Krongold}, {Nicastro}, {Brickhouse},
  {Elvis}, {Liedahl}, \& {Mathur}}]{Krongold2003}
{Krongold}, Y., {Nicastro}, F., {Brickhouse}, N.~S., {et~al.} 2003, \apj, 597,
  832, \dodoi{10.1086/378639}

\bibitem[{{Krongold} {et~al.}(2007){Krongold}, {Nicastro}, {Elvis},
  {Brickhouse}, {Binette}, {Mathur}, \&
  {Jim{\'e}nez-Bail{\'o}n}}]{Krongold2007}
{Krongold}, Y., {Nicastro}, F., {Elvis}, M., {et~al.} 2007, \apj, 659, 1022,
  \dodoi{10.1086/512476}

\bibitem[{{Krongold} {et~al.}(2005){Krongold}, {Nicastro}, {Elvis},
  {Brickhouse}, {Mathur}, \& {Zezas}}]{Krongold2005}
---. 2005, \apj, 620, 165, \dodoi{10.1086/425293}

\bibitem[{{Laha} {et~al.}(2016){Laha}, {Guainazzi}, {Chakravorty}, {Dewangan},
  \& {Kembhavi}}]{Laha2016}
{Laha}, S., {Guainazzi}, M., {Chakravorty}, S., {Dewangan}, G.~C., \&
  {Kembhavi}, A.~K. 2016, \mnras, 457, 3896, \dodoi{10.1093/mnras/stw211}

\bibitem[{{Laha} {et~al.}(2014){Laha}, {Guainazzi}, {Dewangan}, {Chakravorty},
  \& {Kembhavi}}]{Laha2014}
{Laha}, S., {Guainazzi}, M., {Dewangan}, G.~C., {Chakravorty}, S., \&
  {Kembhavi}, A.~K. 2014, \mnras, 441, 2613, \dodoi{10.1093/mnras/stu669}

\bibitem[{{Laha} {et~al.}(2021){Laha}, {Reynolds}, {Reeves}, {Kriss},
  {Guainazzi}, {Smith}, {Veilleux}, \& {Proga}}]{Laha2021}
{Laha}, S., {Reynolds}, C.~S., {Reeves}, J., {et~al.} 2021, Nature Astronomy,
  5, 13, \dodoi{10.1038/s41550-020-01255-2}

\bibitem[{{Lodders} {et~al.}(2009){Lodders}, {Palme}, \& {Gail}}]{Lodders2009}
{Lodders}, K., {Palme}, H., \& {Gail}, H.~P. 2009, Landolt B\&ouml;rnstein, 4B,
  712, \dodoi{10.1007/978-3-540-88055-4_34}

\bibitem[{{Maitra} \& {Miller}(2010)}]{Maitra2010}
{Maitra}, D., \& {Miller}, J.~M. 2010, \apj, 718, 551,
  \dodoi{10.1088/0004-637X/718/1/551}

\bibitem[{{Mao} {et~al.}(2017){Mao}, {Kaastra}, {Mehdipour}, {Raassen}, {Gu},
  \& {Miller}}]{Mao2017}
{Mao}, J., {Kaastra}, J.~S., {Mehdipour}, M., {et~al.} 2017, \aap, 607, A100,
  \dodoi{10.1051/0004-6361/201731378}

\bibitem[{{Mao} {et~al.}(2018){Mao}, {Kaastra}, {Mehdipour}, {Gu},
  {Costantini}, {Kriss}, {Bianchi}, {Branduardi-Raymont}, {Behar}, {Di Gesu},
  {Ponti}, {Petrucci}, \& {Ebrero}}]{Mao2018}
---. 2018, \aap, 612, A18, \dodoi{10.1051/0004-6361/201732162}

\bibitem[{{Mao} {et~al.}(2019){Mao}, {Mehdipour}, {Kaastra}, {Costantini},
  {Pinto}, {Branduardi-Raymont}, {Behar}, {Peretz}, {Bianchi}, {Kriss},
  {Ponti}, {De Marco}, {Petrucci}, {Di Gesu}, {Middei}, {Ebrero}, \&
  {Arav}}]{Mao2019}
{Mao}, J., {Mehdipour}, M., {Kaastra}, J.~S., {et~al.} 2019, \aap, 621, A99,
  \dodoi{10.1051/0004-6361/201833191}

\bibitem[{{Mao} {et~al.}(2022){Mao}, {Kaastra}, {Mehdipour}, {Kriss}, {Wang},
  {Grafton-Waters}, {Branduardi-Raymont}, {Pinto}, {Landt}, {Walton},
  {Costantini}, {Di Gesu}, {Bianchi}, {Petrucci}, {De Marco}, {Ponti},
  {Fukazawa}, {Ebrero}, \& {Behar}}]{Mao2022}
{Mao}, J., {Kaastra}, J.~S., {Mehdipour}, M., {et~al.} 2022, \aap, 665, A72,
  \dodoi{10.1051/0004-6361/202142637}

\bibitem[{{Matzeu} {et~al.}(2017){Matzeu}, {Reeves}, {Braito}, {Nardini},
  {McLaughlin}, {Lobban}, {Tombesi}, \& {Costa}}]{Matzeu2017}
{Matzeu}, G.~A., {Reeves}, J.~N., {Braito}, V., {et~al.} 2017, \mnras, 472,
  L15, \dodoi{10.1093/mnrasl/slx129}

\bibitem[{{McKernan} {et~al.}(2007){McKernan}, {Yaqoob}, \&
  {Reynolds}}]{McKernan2007}
{McKernan}, B., {Yaqoob}, T., \& {Reynolds}, C.~S. 2007, \mnras, 379, 1359,
  \dodoi{10.1111/j.1365-2966.2007.11993.x}

\bibitem[{{Mehdipour} {et~al.}(2016){Mehdipour}, {Kaastra}, \&
  {Kallman}}]{Mehdipour2016}
{Mehdipour}, M., {Kaastra}, J.~S., \& {Kallman}, T. 2016, \aap, 596, A65,
  \dodoi{10.1051/0004-6361/201628721}

\bibitem[{{Mehdipour} {et~al.}(2015){Mehdipour}, {Kaastra}, {Kriss}, {Cappi},
  {Petrucci}, {Steenbrugge}, {Arav}, {Behar}, {Bianchi}, {Boissay},
  {Branduardi-Raymont}, {Costantini}, {Ebrero}, {Di Gesu}, {Harrison}, {Kaspi},
  {De Marco}, {Matt}, {Paltani}, {Peterson}, {Ponti}, {Pozo Nu{\~n}ez}, {De
  Rosa}, {Ursini}, {de Vries}, {Walton}, \& {Whewell}}]{Mehdipour2015}
{Mehdipour}, M., {Kaastra}, J.~S., {Kriss}, G.~A., {et~al.} 2015, \aap, 575,
  A22, \dodoi{10.1051/0004-6361/201425373}

\bibitem[{{Middleton} \& {Done}(2010)}]{Middleton2010}
{Middleton}, M., \& {Done}, C. 2010, \mnras, 403, 9,
  \dodoi{10.1111/j.1365-2966.2009.15969.x}

\bibitem[{{Middleton} {et~al.}(2009){Middleton}, {Done}, {Ward},
  {Gierli{\'n}ski}, \& {Schurch}}]{Middleton2009}
{Middleton}, M., {Done}, C., {Ward}, M., {Gierli{\'n}ski}, M., \& {Schurch}, N.
  2009, \mnras, 394, 250, \dodoi{10.1111/j.1365-2966.2008.14255.x}

\bibitem[{{Middleton} {et~al.}(2011){Middleton}, {Uttley}, \&
  {Done}}]{Middleton2011}
{Middleton}, M., {Uttley}, P., \& {Done}, C. 2011, \mnras, 417, 250,
  \dodoi{10.1111/j.1365-2966.2011.19185.x}

\bibitem[{{Miller} {et~al.}(2015){Miller}, {Kaastra}, {Miller}, {Reynolds},
  {Brown}, {Cenko}, {Drake}, {Gezari}, {Guillochon}, {Gultekin}, {Irwin},
  {Levan}, {Maitra}, {Maksym}, {Mushotzky}, {O'Brien}, {Paerels}, {de Plaa},
  {Ramirez-Ruiz}, {Strohmayer}, \& {Tanvir}}]{Miller2015}
{Miller}, J.~M., {Kaastra}, J.~S., {Miller}, M.~C., {et~al.} 2015, \nat, 526,
  542, \dodoi{10.1038/nature15708}

\bibitem[{{Mitsuda} {et~al.}(1984){Mitsuda}, {Inoue}, {Koyama}, {Makishima},
  {Matsuoka}, {Ogawara}, {Shibazaki}, {Suzuki}, {Tanaka}, \&
  {Hirano}}]{Mitsuda1984}
{Mitsuda}, K., {Inoue}, H., {Koyama}, K., {et~al.} 1984, \pasj, 36, 741

\bibitem[{{Mizumoto} {et~al.}(2019){Mizumoto}, {Done}, {Tomaru}, \&
  {Edwards}}]{Mizumoto2019}
{Mizumoto}, M., {Done}, C., {Tomaru}, R., \& {Edwards}, I. 2019, \mnras, 489,
  1152, \dodoi{10.1093/mnras/stz2225}

\bibitem[{{Nenkova} {et~al.}(2008){Nenkova}, {Sirocky}, {Nikutta},
  {Ivezi{\'c}}, \& {Elitzur}}]{Nenkova2008}
{Nenkova}, M., {Sirocky}, M.~M., {Nikutta}, R., {Ivezi{\'c}}, {\v{Z}}., \&
  {Elitzur}, M. 2008, \apj, 685, 160, \dodoi{10.1086/590483}

\bibitem[{{Netzer} {et~al.}(2004){Netzer}, {Shemmer}, {Maiolino}, {Oliva},
  {Croom}, {Corbett}, \& {di Fabrizio}}]{Netzer2004}
{Netzer}, H., {Shemmer}, O., {Maiolino}, R., {et~al.} 2004, \apj, 614, 558,
  \dodoi{10.1086/423608}

\bibitem[{{Nicastro} {et~al.}(1999){Nicastro}, {Fiore}, \&
  {Matt}}]{Nicastro1999}
{Nicastro}, F., {Fiore}, F., \& {Matt}, G. 1999, \apj, 517, 108,
  \dodoi{10.1086/307187}

\bibitem[{{Nicastro} {et~al.}(2016){Nicastro}, {Senatore}, {Krongold},
  {Mathur}, \& {Elvis}}]{Nicastro2016}
{Nicastro}, F., {Senatore}, F., {Krongold}, Y., {Mathur}, S., \& {Elvis}, M.
  2016, \apjl, 828, L12, \dodoi{10.3847/2041-8205/828/1/L12}

\bibitem[{{Parker} {et~al.}(2017){Parker}, {Pinto}, {Fabian}, {Lohfink},
  {Buisson}, {Alston}, {Kara}, {Cackett}, {Chiang}, {Dauser}, {De Marco},
  {Gallo}, {Garcia}, {Harrison}, {King}, {Middleton}, {Miller}, {Miniutti},
  {Reynolds}, {Uttley}, {Vasudevan}, {Walton}, {Wilkins}, \&
  {Zoghbi}}]{Parker2017}
{Parker}, M.~L., {Pinto}, C., {Fabian}, A.~C., {et~al.} 2017, \nat, 543, 83,
  \dodoi{10.1038/nature21385}

\bibitem[{{Pinto} {et~al.}(2013){Pinto}, {Kaastra}, {Costantini}, \& {de
  Vries}}]{Pinto2013}
{Pinto}, C., {Kaastra}, J.~S., {Costantini}, E., \& {de Vries}, C. 2013, \aap,
  551, A25, \dodoi{10.1051/0004-6361/201220481}

\bibitem[{{Proga} \& {Kallman}(2004)}]{Proga2004}
{Proga}, D., \& {Kallman}, T.~R. 2004, \apj, 616, 688, \dodoi{10.1086/425117}

\bibitem[{{Qu} {et~al.}(2020){Qu}, {Bregman}, {Hodges-Kluck}, {Li}, \&
  {Lindley}}]{Qu2020}
{Qu}, Z., {Bregman}, J.~N., {Hodges-Kluck}, E., {Li}, J.-T., \& {Lindley}, R.
  2020, \apj, 894, 142, \dodoi{10.3847/1538-4357/ab774e}

\bibitem[{{Reeves} \& {Braito}(2019)}]{Reeves2019}
{Reeves}, J.~N., \& {Braito}, V. 2019, \apj, 884, 80,
  \dodoi{10.3847/1538-4357/ab41f9}

\bibitem[{{Reeves} {et~al.}(2013){Reeves}, {Porquet}, {Braito}, {Gofford},
  {Nardini}, {Turner}, {Crenshaw}, \& {Kraemer}}]{Reeves2013}
{Reeves}, J.~N., {Porquet}, D., {Braito}, V., {et~al.} 2013, \apj, 776, 99,
  \dodoi{10.1088/0004-637X/776/2/99}

\bibitem[{{Reeves} {et~al.}(2009){Reeves}, {Sambruna}, {Braito}, \&
  {Eracleous}}]{Reeves2009}
{Reeves}, J.~N., {Sambruna}, R.~M., {Braito}, V., \& {Eracleous}, M. 2009,
  \apjl, 702, L187, \dodoi{10.1088/0004-637X/702/2/L187}

\bibitem[{{Reynolds}(1997)}]{Reynolds1997}
{Reynolds}, C.~S. 1997, \mnras, 286, 513, \dodoi{10.1093/mnras/286.3.513}

\bibitem[{{Rogantini} {et~al.}(2022){Rogantini}, {Mehdipour}, {Kaastra},
  {Costantini}, {Jur{\'a}{\v{n}}ov{\'a}}, \& {Kara}}]{Rogantini2022}
{Rogantini}, D., {Mehdipour}, M., {Kaastra}, J., {et~al.} 2022, \apj, 940, 122,
  \dodoi{10.3847/1538-4357/ac9c01}

\bibitem[{{Rybicki} \& {Lightman}(1986)}]{Rybicki1986}
{Rybicki}, G.~B., \& {Lightman}, A.~P. 1986, {Radiative Processes in
  Astrophysics}

\bibitem[{{Sako} {et~al.}(2001){Sako}, {Kahn}, {Behar}, {Kaastra}, {Brinkman},
  {Boller}, {Puchnarewicz}, {Starling}, {Liedahl}, {Clavel}, \&
  {Santos-Lleo}}]{Sako2001}
{Sako}, M., {Kahn}, S.~M., {Behar}, E., {et~al.} 2001, \aap, 365, L168,
  \dodoi{10.1051/0004-6361:20000081}

\bibitem[{{Savage} \& {Wakker}(2009)}]{Savage2009}
{Savage}, B.~D., \& {Wakker}, B.~P. 2009, \apj, 702, 1472,
  \dodoi{10.1088/0004-637X/702/2/1472}

\bibitem[{{Sembach} {et~al.}(2003){Sembach}, {Wakker}, {Savage}, {Richter},
  {Meade}, {Shull}, {Jenkins}, {Sonneborn}, \& {Moos}}]{Sembach2003}
{Sembach}, K.~R., {Wakker}, B.~P., {Savage}, B.~D., {et~al.} 2003, \apjs, 146,
  165, \dodoi{10.1086/346231}

\bibitem[{{Shakura} \& {Sunyaev}(1973)}]{Shakura1973}
{Shakura}, N.~I., \& {Sunyaev}, R.~A. 1973, \aap, 24, 337

\bibitem[{{Silva} {et~al.}(2016){Silva}, {Uttley}, \& {Costantini}}]{Silva2016}
{Silva}, C.~V., {Uttley}, P., \& {Costantini}, E. 2016, \aap, 596, A79,
  \dodoi{10.1051/0004-6361/201628555}

\bibitem[{{Steenbrugge} {et~al.}(2003){Steenbrugge}, {Kaastra}, {de Vries}, \&
  {Edelson}}]{Steenbrugge2003}
{Steenbrugge}, K.~C., {Kaastra}, J.~S., {de Vries}, C.~P., \& {Edelson}, R.
  2003, \aap, 402, 477, \dodoi{10.1051/0004-6361:20030261}

\bibitem[{{Steenbrugge} {et~al.}(2005){Steenbrugge}, {Kaastra}, {Crenshaw},
  {Kraemer}, {Arav}, {George}, {Liedahl}, {van der Meer}, {Paerels}, {Turner},
  \& {Yaqoob}}]{Steenbrugge2005}
{Steenbrugge}, K.~C., {Kaastra}, J.~S., {Crenshaw}, D.~M., {et~al.} 2005, \aap,
  434, 569, \dodoi{10.1051/0004-6361:20047138}

\bibitem[{{Tarter} {et~al.}(1969){Tarter}, {Tucker}, \&
  {Salpeter}}]{Tarter1969}
{Tarter}, C.~B., {Tucker}, W.~H., \& {Salpeter}, E.~E. 1969, \apj, 156, 943,
  \dodoi{10.1086/150026}

\bibitem[{{Tombesi} {et~al.}(2013){Tombesi}, {Cappi}, {Reeves}, {Nemmen},
  {Braito}, {Gaspari}, \& {Reynolds}}]{Tombesi2013}
{Tombesi}, F., {Cappi}, M., {Reeves}, J.~N., {et~al.} 2013, \mnras, 430, 1102,
  \dodoi{10.1093/mnras/sts692}

\bibitem[{{Tumlinson} {et~al.}(2017){Tumlinson}, {Peeples}, \&
  {Werk}}]{Tumlinson2017}
{Tumlinson}, J., {Peeples}, M.~S., \& {Werk}, J.~K. 2017, \araa, 55, 389,
  \dodoi{10.1146/annurev-astro-091916-055240}

\bibitem[{{Vaughan}(2005)}]{Vaughan2005}
{Vaughan}, S. 2005, \aap, 431, 391, \dodoi{10.1051/0004-6361:20041453}

\bibitem[{{Vaughan} {et~al.}(2003){Vaughan}, {Edelson}, {Warwick}, \&
  {Uttley}}]{Vaughan2003}
{Vaughan}, S., {Edelson}, R., {Warwick}, R.~S., \& {Uttley}, P. 2003, \mnras,
  345, 1271, \dodoi{10.1046/j.1365-2966.2003.07042.x}

\bibitem[{{Wang} {et~al.}(2022{\natexlab{a}}){Wang}, {He}, {Mao}, {Kaastra},
  {Xue}, \& {Mehdipour}}]{Wang2022ApJ}
{Wang}, Y., {He}, Z., {Mao}, J., {et~al.} 2022{\natexlab{a}}, \apj, 928, 7,
  \dodoi{10.3847/1538-4357/ac524d}

\bibitem[{{Wang} {et~al.}(2022{\natexlab{b}}){Wang}, {Kaastra}, {Mehdipour},
  {Mao}, {Costantini}, {Kriss}, {Pinto}, {Ponti}, {Behar}, {Bianchi},
  {Branduardi-Raymont}, {De Marco}, {Grafton-Waters}, {Petrucci}, {Ebrero},
  {Walton}, {Kaspi}, {Xue}, {Paltani}, {di Gesu}, \& {He}}]{Wang2022}
{Wang}, Y., {Kaastra}, J., {Mehdipour}, M., {et~al.} 2022{\natexlab{b}}, \aap,
  657, A77, \dodoi{10.1051/0004-6361/202141599}

\bibitem[{{Xia} {et~al.}(2024){Xia}, {Liu}, \& {Xue}}]{Xia2024}
{Xia}, R., {Liu}, H., \& {Xue}, Y. 2024, \apjl, 961, L32,
  \dodoi{10.3847/2041-8213/ad1bf2}

\bibitem[{{Zoghbi} \& {Fabian}(2011)}]{Zoghbi2011}
{Zoghbi}, A., \& {Fabian}, A.~C. 2011, \mnras, 418, 2642,
  \dodoi{10.1111/j.1365-2966.2011.19655.x}

\end{thebibliography}
\bibliographystyle{aasjournal}

\appendix

\section{$1-4$~keV Light Curve and Power Spectral Distribution} \label{sec:LC_PSD}

\begin{figure}[!ht]
\centering
\includegraphics[width=0.85\textwidth]{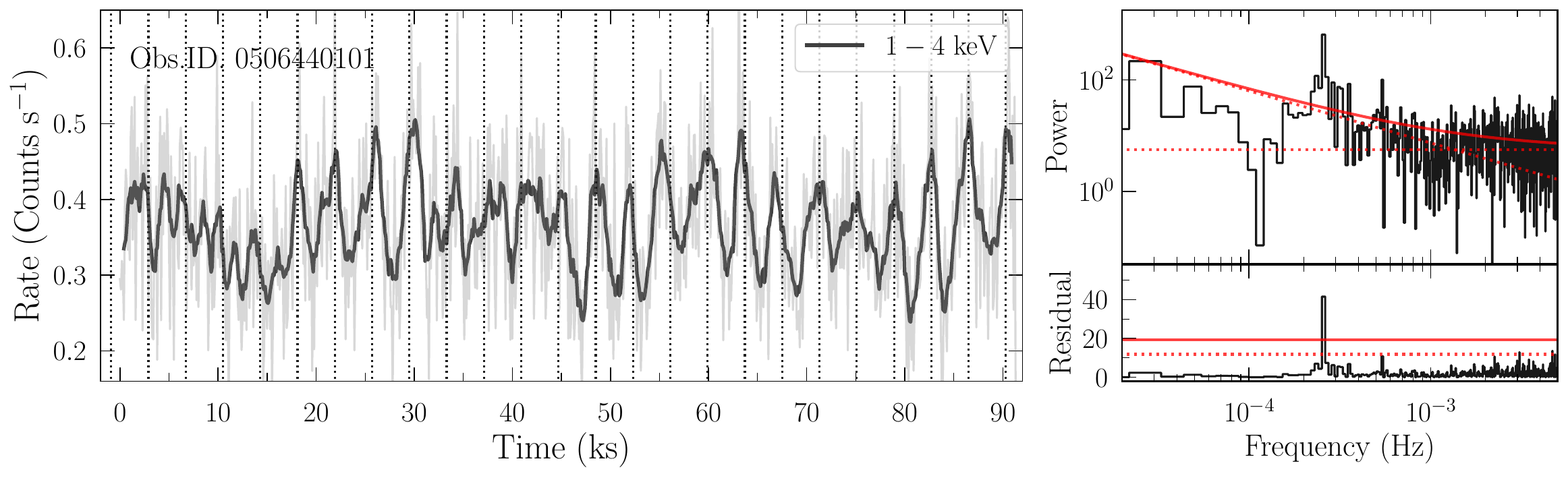} \\
\includegraphics[width=0.85\textwidth]{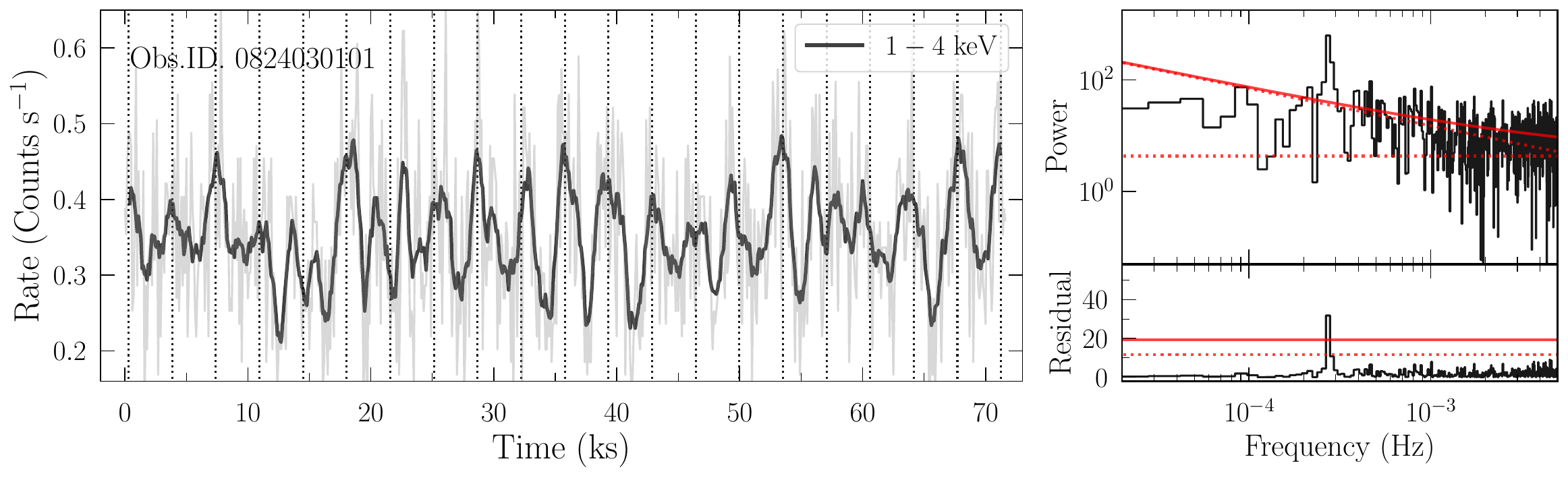} 
\caption{$1-4$~keV light curve (left) and power spectral distribution (right) for the May-2007 (top) and Oct-2018 (bottom) observations, respectively. The gray curve in the light-curve panel is the original light curve in a bin size of $100$~s, with the black curve showing the nine-point moving-averaged light curve. The separation of vertical dotted lines is equal to the best-fit QPO period. In the PSD panel, spectral power is calculated in the unit of [rms/mean$^2$~Hz$^{-1}$]. The solid red curve is the best-fit continuum model, with the dotted curves its decomposition. Residual is defined by $2\times$Data/Model. The horizon solid and dotted lines mark the $4\,\sigma$ and $3\,\sigma$ QPO detectability thresholds, respectively.
}
\label{fig:LC_PSD}
\end{figure}

We extracted the $1-4$~keV EPIC-pn light curves (LCs) of the May-2007 and Oct-2018 observations for illustration purposes of the $\sim2.7\times10^{-4}$~Hz QPO. Detailed QPO analysis of observations before 2018 can be found in \cite{Jin2020}, and see \cite{Xia2024} for the 2020-2021 observations. The SAS task {\it epiclccorr} was adopted to generate background-corrected LCs. We adopted the {\it Powerspectrum} function included in the spectral timing package {\it Stingray} \citep{Huppenkothen2019} to calculate the power spectral distribution (PSD). We fit the PSD continuum using a simple but popular model, which consists of a power law for the red (low-frequency) noise and a constant for the Poisson (high-frequency) noise (e.g., \citealp{Vaughan2003, Vaughan2005, Alston2014}). The maximum-likelihood estimation was applied in the continuum fitting, using the Whittle likelihood function of $\log p = (I | \theta, H) = - \Sigma_j \frac{I_j}{S_j} + \log S_j$. In the equation, $I_j$ and $S_j$ are the $j$-th points of the observed PSD and the continuum model, respectively. $H$ denotes the applied continuum model, with $\theta$ the fitting parameters. The fitting residual is defined by $2\times$Data/Model. The QPO significance was estimated by comparing the residual with the $\chi^2$ distribution with two degrees of freedom. This method serves as a good approximation to the frequently adopted method using Markov Chain Monte Carlo sampling \citep{Jin2020}. Figure~\ref{fig:LC_PSD} shows the derived LCs and PSDs.

\section{Searching for Ultra-fast Outflow}
\label{sec:ufo_search}

\begin{figure}
\centering
\includegraphics[width=0.47\textwidth]{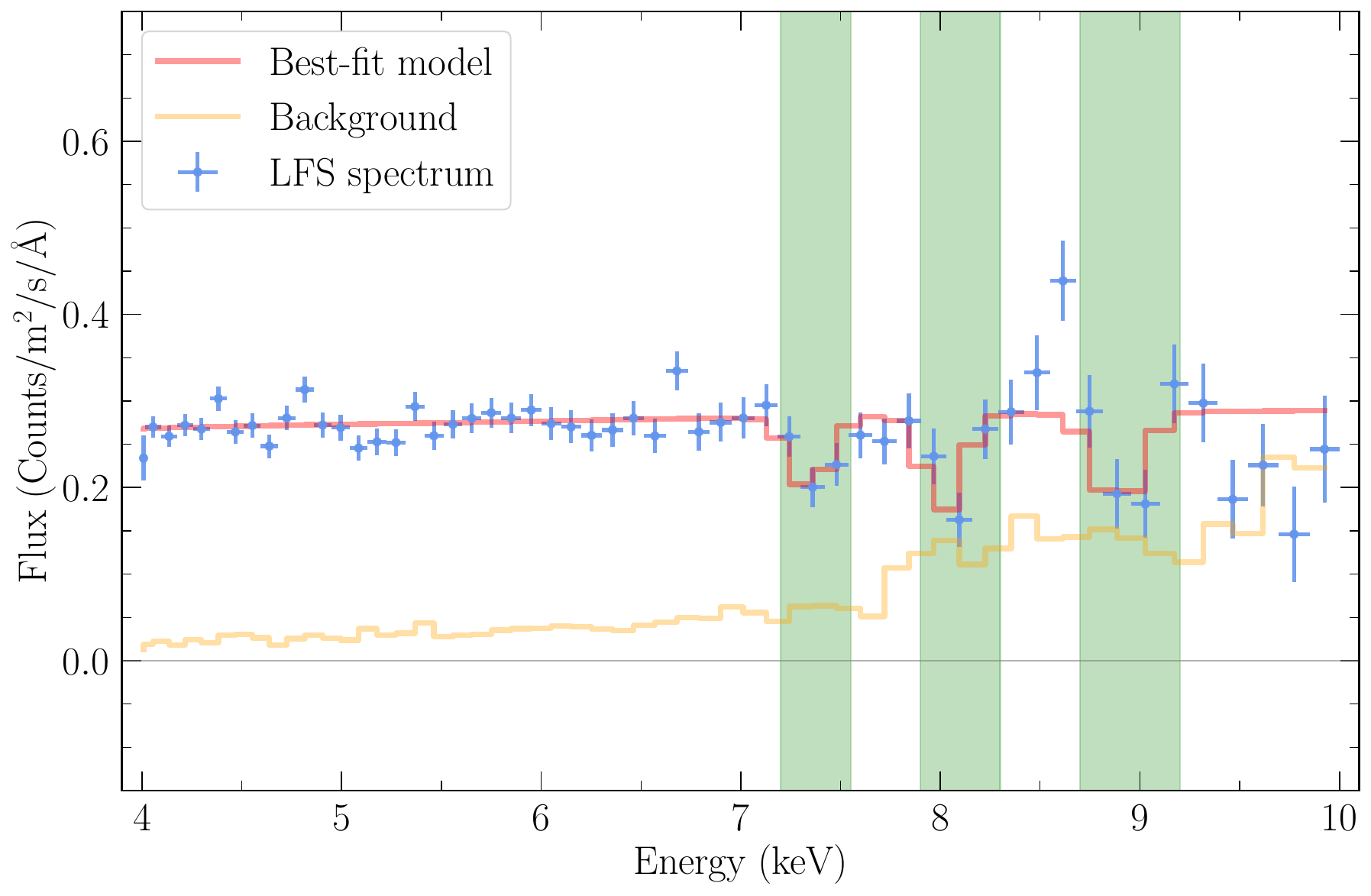} 
\includegraphics[width=0.47\textwidth]{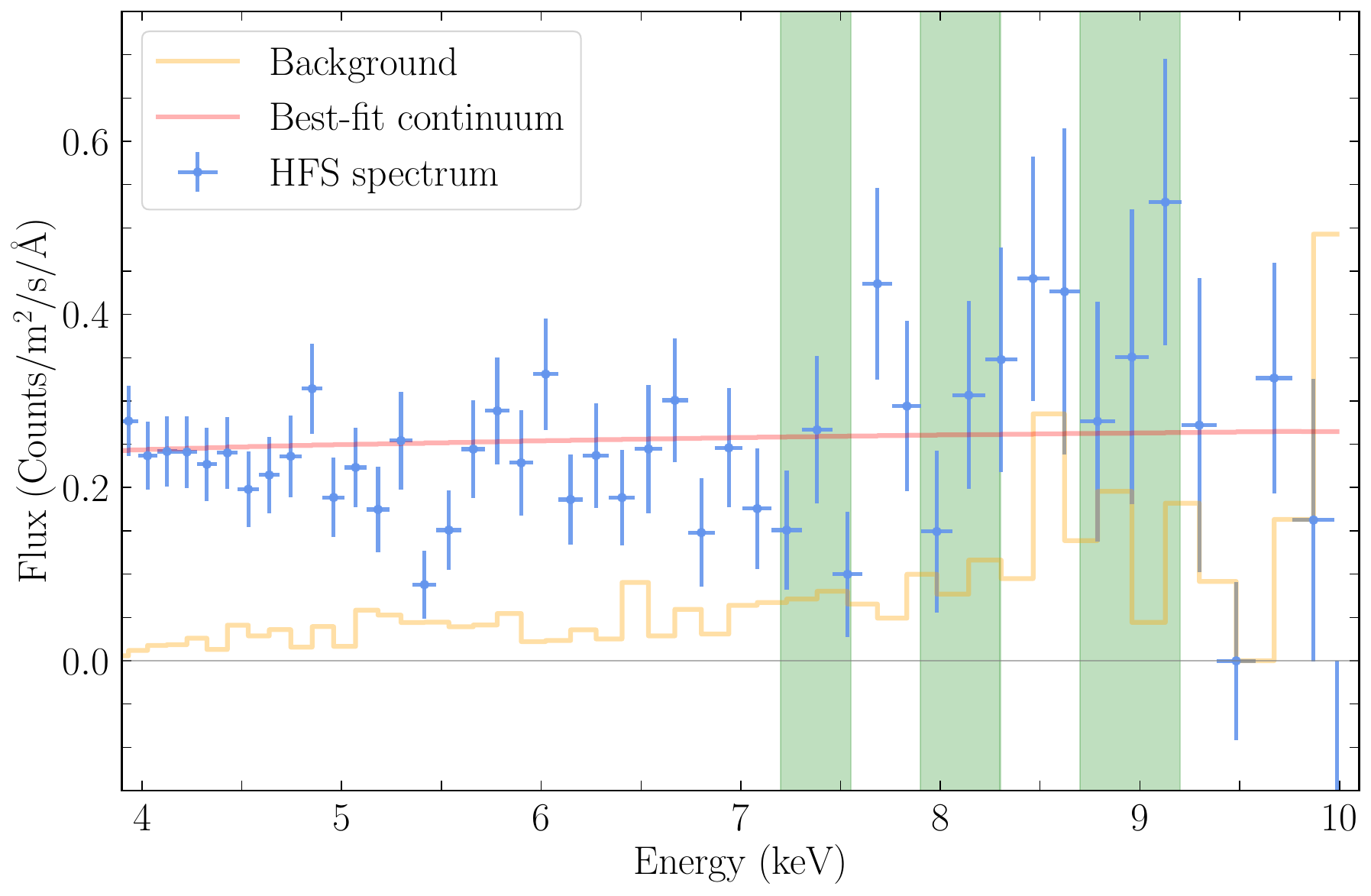} 
\caption{Stacked $4-10$~keV EPIC-pn spectra of the LFS (left) and HFS (right), respectively. Blue points are the observed data. The orange curve shows the background level. We fit the spectral continuum by a power law. For the LFS spectrum, we also use three Gaussian profiles to fit the potential UFO features. The red curve is the best-fit model. The vertical green shadows in both panels highlight the potential UFO features seen in the stacked LFS spectrum.
}
\label{fig:stacked_4-10}
\end{figure}

Targets with a high Eddington ratio are usually accompanied by ultra-fast outflow (e.g., \citealp{Matzeu2017, Parker2017, Reeves2019}). Figure~\ref{fig:stacked_4-10} shows the stacked spectra of the LFS and HFS within the $4-10$~keV range. We discovered three potential absorption features associated with the UFO at $\sim7.4$~keV, $\sim8.1$~keV, and $\sim9.0$~keV in the stacked LFS spectrum. We adopted a power law to fit the spectral continuum and used three Gaussian profiles to fit the absorption lines. Table~\ref{tab:UFO_detection} lists the best-fit parameters of the absorption lines. We estimated the line significance using the corrected Akaike information criterion ($\Delta{\rm AIC_c}$) between models with and without the interested line \citep{Burnham2011}. Two of the lines show moderate significance at $2\,\sigma$ level, while the third is considered insignificant. We calculated the line velocity shift assuming an origin of Fe\,{\sc xxv} He$\alpha$ or Fe\,{\sc xxvi} Ly$\alpha$. The calculating result shows no overlapping of the line velocities, suggesting the three lines do not come from the same UFO component. All three lines are insignificant in the stacked HFS spectrum. The UFO is implied but cannot be confirmed by the current analysis.

\renewcommand\arraystretch{1.1}
\begin{deluxetable}{ccccccccc}
\tablewidth{0pt}
\setlength{\tabcolsep}{3. mm}
\tablecaption{Properties of the potential UFO lines in the stacked LFS spectrum. Columns are (1) observed line energy, (2) energy at the AGN rest frame, (3) full width at half maximum, (4) equivalent width, (5)-(6) outflow velocity if Fe\,{\sc xxv} He$\alpha$ line or Fe\,{\sc xxvi} Ly$\alpha$ line, (7)-(8) $\Delta C_{\rm stat}$ and $\Delta {\rm AIC_c}$ with and without the line, and (9) signal-to-noise ratio of the line. 
\label{tab:UFO_detection}}
\tablewidth{400pt}
\tablehead{ 
\colhead{$E_{\rm obs}$} & \colhead{$E_{\rm rest}$} & \colhead{$FWHM$} & \colhead{EW} & \colhead{$v$ (Fe~{\sc xxv} He$\alpha$)} & \colhead{$v$ (Fe~{\sc xxvi} Ly$\alpha$)} & \colhead{$\Delta C_{\rm stat}$} & \colhead{$\Delta {\rm AIC_c}$} & \colhead{$SNR$} \\
\colhead{(keV)} & \colhead{(keV)} & \colhead{(keV)} & \colhead{(eV)} & \colhead{($c$)} & \colhead{($c$)} & \colhead{} & \colhead{} 
}
\decimalcolnumbers
\startdata
$7.39^{+0.04}_{-0.03}$ & $7.71^{+0.04}_{-0.03}$ &  $0.10^{+0.18}_{-0.08}$ & $77.8$ & $-0.131^{+0.004}_{-0.003}$ & $-0.096^{+0.005}_{-0.004}$ & $12.9$ & $5.9$ & $1.9\,\sigma$  \\
$8.07^{+0.03}_{-0.02}$ & $8.42^{+0.03}_{-0.02}$ &  $0.03^{+0.11}_{-0.01}$ & $98.4$ & $-0.205^{+0.003}_{-0.002}$ & $-0.173^{+0.003}_{-0.002}$ & $13.0$ & $6.0$ & $2.0\,\sigma$ \\
$8.95^{+0.03}_{-0.03}$ & $9.34^{+0.03}_{-0.03}$ &  $0.04^{+0.11}_{-0.02}$ & $110.6$ & $-0.283^{+0.002}_{-0.002}$ & $-0.254^{+0.002}_{-0.002}$  & $8.6$ & $1.6$ & $<1.0\,\sigma$ \\
\enddata
\end{deluxetable}

\section{Modeling the BEB as A Photoionized Plasma}
\label{sec:BEB_pion}

\begin{figure}
\centering
\includegraphics[width=0.47\textwidth]{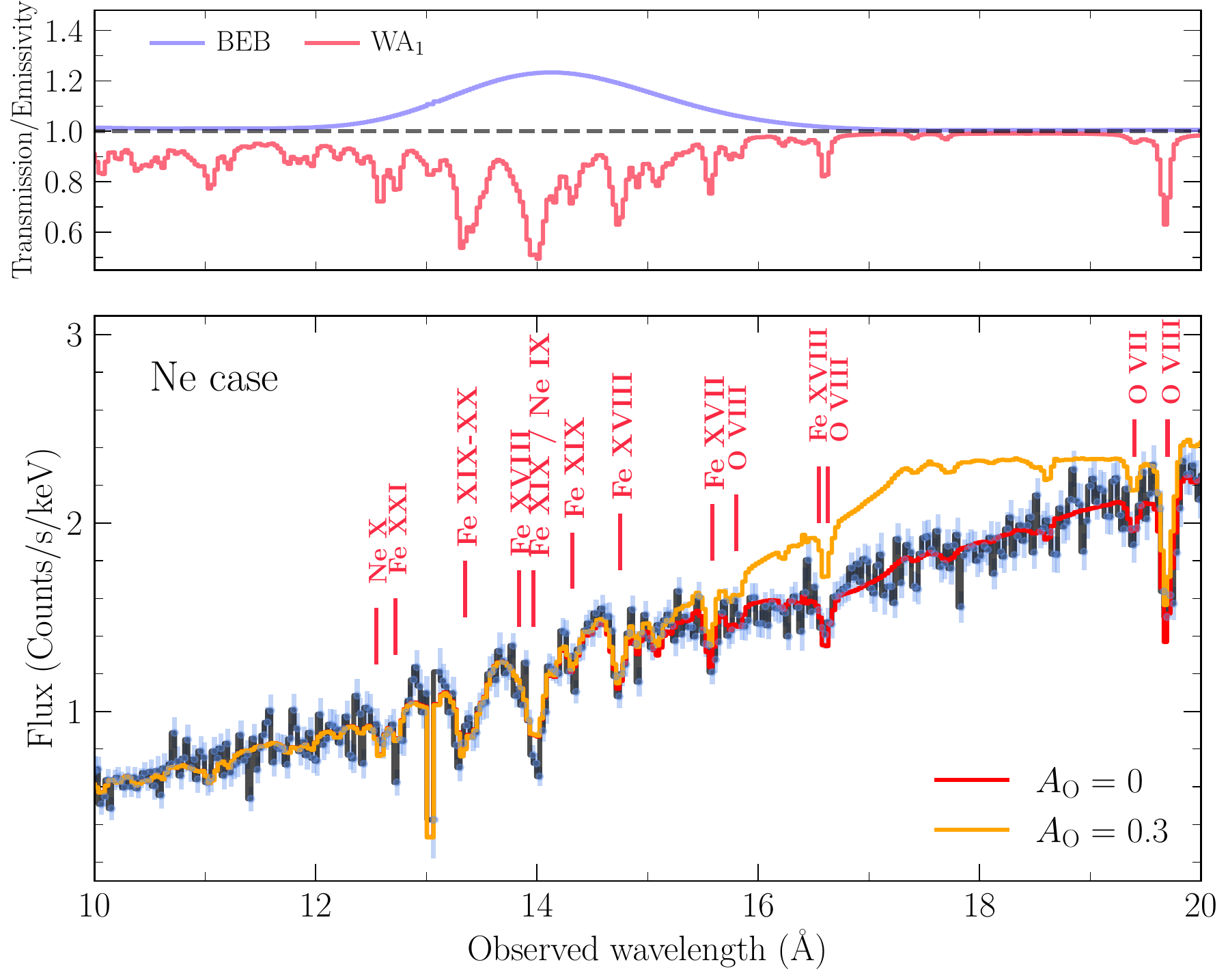} 
\includegraphics[width=0.47\textwidth]{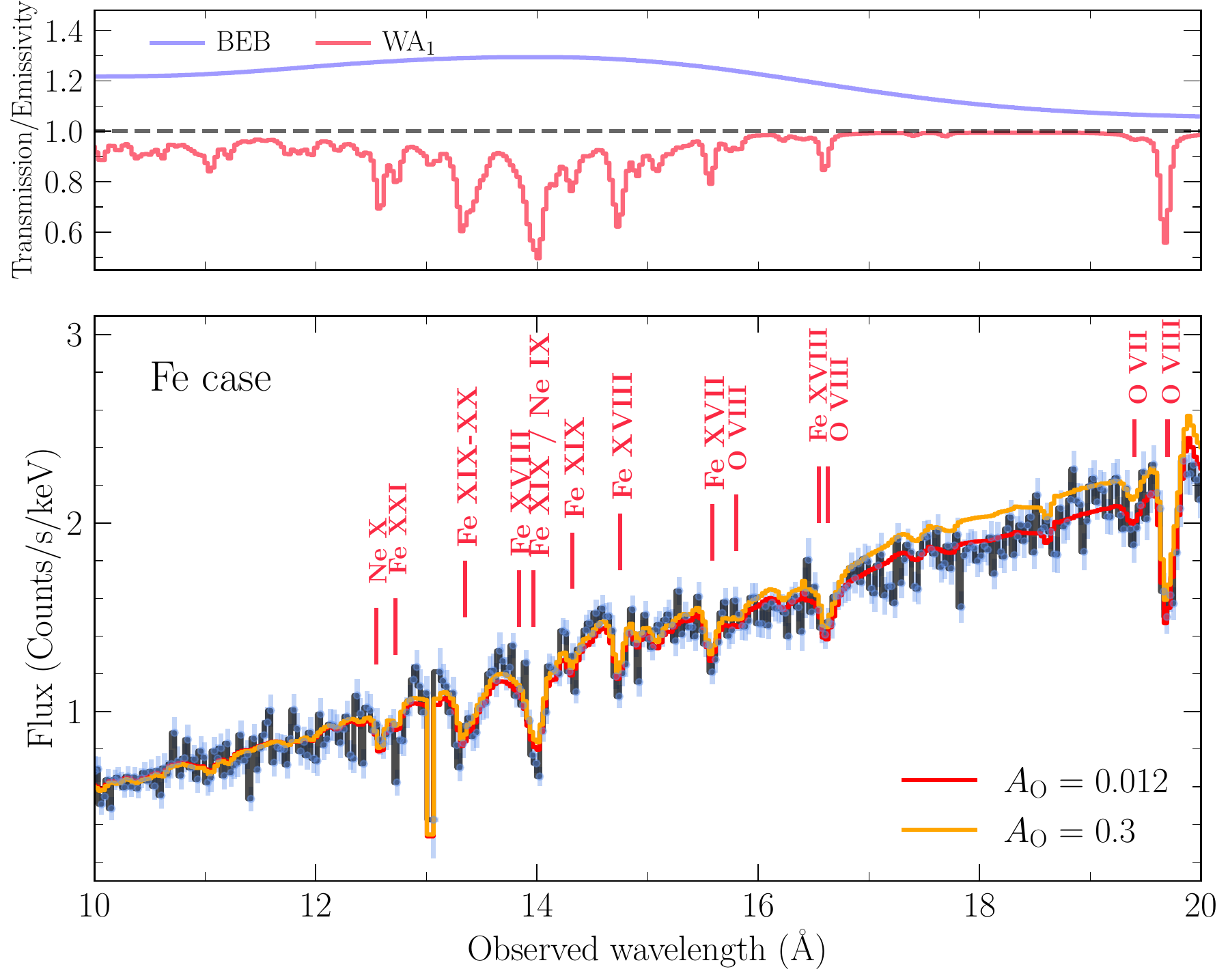} 
\caption{Stacked LFS BEB spectra fitted by the relativistically broadened Ne (left) or Fe (right) lines. The top panel only shows the decomposition of BEB and ${\rm WA}_1$ for clarity. The red curve in the spectrum panel is the best-fit model. The orange curve is the best model but with $0.3$ times the solar oxygen abundance.
}
\label{fig:BEB_pion_spec}
\end{figure}

Centered at around $14$~\AA, the BEB may be relativistically broadened Ne or Fe lines, which possess strong features at this wavelength band. Based on the best-fit stacked LFS spectrum, we replaced the Gaussian profile with a {\it pion} emission component to model the BEB. When examining the Ne case, we fixed the Ne abundance at the solar value but let C, N, O, and Fe abundances free to vary. As for the Fe case, we fixed the Fe abundance to solar value, and the abundances of C, N, O, Ne, Mg, Al, Si, and S were treated as free parameters. The best-fit $10-18$~\AA\ spectra are displayed in Figure~\ref{fig:BEB_pion_spec}.

The best-fit Ne case results in a fitting statistic of $C_{\rm stat} / {\rm DoF} = 1470/1026$. This explanation requires an ionization state of $\log (\xi/{\rm erg~cm~s}^{-1}) \sim 1.5$ with a velocity of $\sim3000$~km~s$^{-1}$ (inflow) and a broadening of $\sim20000$~km~s$^{-1}$. The best-fit abundances of O and Fe are near zero, avoiding the generation of strong emission features unseen in the observed spectrum (see the orange curve in Figure~\ref{fig:BEB_pion_spec} for an example). The best C and N abundances are around $0.1$ solar.

The Fe case achieves a slightly better statistic of $C_{\rm stat} / {\rm DoF} = 1457/1022$. In this case, the ionization state is similar to the high-ionization WA value of $\log (\xi/{\rm erg~cm~s}^{-1}) \sim 4.0$. This explanation requires a velocity of $\sim-9000$~km~s$^{-1}$ (outflow) with a broadening of $\sim30000$~km~s$^{-1}$. The best-fit abundances of N, O, Ne, Mg, Al, and S are near zero ($0.012$ solar abundance for oxygen), and C and Si abundances are $0.65$ and $0.5$ solar, respectively. The impact of oxygen abundance is less pronounced in this case, and setting the oxygen abundance to $0.3$ solar only marginally increases the continuum level at around $19$~\AA. However, the best-fit oxygen abundance of $0.012$ solar indicates that the Fe abundance is $83$ times the oxygen abundance.

The weird abundance composition is unusual and has never been discovered before for the photoionized plasma. Moreover, the best-fit statistics of the two cases are worse than the original model using a simple Gaussian profile ($C_{\rm stat} / {\rm DoF} = 1461/1032$). We did not find a photoionized explanation using a narrow component (i.e., multiple narrow Fe emission lines). Currently, the BEB nature remains uncertain and requires future investigations.




\end{CJK*}
\end{document}